\documentclass[useAMS,usenatbib]{mn2e}

\usepackage[english]{babel}
\usepackage{amsmath,amssymb}
\usepackage{graphicx, subfigure}
\usepackage[normalem]{ulem}
\usepackage{color}

\def\be{\begin{equation}} 
\def\ee{\end{equation}} 
\def\ba{\begin{eqnarray}} 
\def\ea{\end{eqnarray}}

\def\HI{\hbox{H~$\scriptstyle\rm I\ $}} 
\def\HII{\hbox{H~$\scriptstyle\rm II\ $}} 
 
\def\HeII{\hbox{He~$\scriptstyle\rm II\ $}} 
 
\def\CIV{\hbox{C~$\scriptstyle\rm IV\ $}} 
 
\def\OVI{\hbox{O~$\scriptstyle\rm VI\ $}}

\def\gsim{\lower.5ex\hbox{\gtsima}} 
\def\lsim{\lower.5ex\hbox{\ltsima}} \def\gtsima{$\; \buildrel > \over 
\sim \;$} \def\ltsima{$\; \buildrel < \over \sim \;$} \def\prosima{$\; 
\buildrel \propto \over \sim \;$} \def\gsim{\lower.5ex\hbox{\gtsima}} 
\def\lsim{\lower.5ex\hbox{\ltsima}} 
\def\simgt{\lower.5ex\hbox{\gtsima}} 
\def\simlt{\lower.5ex\hbox{\ltsima}} 
\def\simpr{\lower.5ex\hbox{\prosima}} \def\la{\lsim} \def\ga{\gsim} 
  
 \def\gtsima{$\; \buildrel > \over \sim \;$} 
\def\ltsima{$\; \buildrel < \over \sim \;$} 
\def\gsim{\lower.5ex\hbox{\gtsima}} 
\def\lsim{\lower.5ex\hbox{\ltsima}} 
\def\simgt{\lower.5ex\hbox{\gtsima}} 
\def\simlt{\lower.5ex\hbox{\ltsima}} 
\def\simpr{\lower.5ex\hbox{\prosima}} 
\def\la{\lsim} 
\def\ga{\gsim}

\def\Lya{Ly$\alpha$~}

\def\E3{{\cal E}_{\rm g}^{III}}

\def\r12{r_{1/2}} 
\def\x12{x_{1/2}} 
\def\v12{v_{1/2}}

\newcommand{\obs}{{\rm obs}}
\newcommand{\equ}{{\rm equ}}
\newcommand{\eff}{{\rm eff}}
\newcommand{\refr}{{\rm ref}}

\newcommand{\kpc}{{\rm kpc}}

\newcommand{\los}{{\rm los}}
\newcommand{\ext}{{\rm ext}}
\newcommand{\tot}{{\rm tot}}
\newcommand{\pdf}{{\rm pdf}}

\newcommand{\frw}{{\rm frw}}

\usepackage{ifpdf}

\title[Simulating intergalactic quasar scintillation]{Simulating intergalactic quasar scintillation}

\author[A. Pallottini, A. Ferrara, C. Evoli]{
A. Pallottini$^{1}$\thanks{E-mail:andrea.pallottini@sns.it},
A. Ferrara$^{1}$ and C. Evoli$^{2}$\\
$^{1}$Scuola Normale Superiore, Piazza dei Cavalieri 7, 56126 Pisa, Italy\\
$^{2}$II. Institut f\"ur Theoretische Physik, Universit\"at Hamburg, Luruper Chaussee 149, D-22761 Hamburg, Germany\\
}

\begin{document}

\date{}

\pagerange{\pageref{firstpage}--\pageref{lastpage}} \pubyear{2013}

\maketitle

\label{firstpage}

\begin{abstract}
Intergalactic scintillation of distant quasars is sensitive to free electrons and therefore complements Ly$\alpha$ absorption line experiments probing the neutral intergalactic medium (IGM). We present a new scheme to compute IGM refractive scintillation effects on distant sources in combination with Adaptive Mesh Refinement cosmological simulations. First we validate our model by reproducing the well-known interstellar scintillation (ISS) of Galactic sources. The simulated cosmic density field is then used to infer the statistical properties of intergalactic scintillation. Contrary to previous claims, we find that the scattering measure of the simulated IGM at $z<2$ is $\langle \mbox{SM}_{\equ}\rangle=3.879$, i.e. almost 40 times larger than for the usually assumed smooth IGM. This yield an average modulation index ranging from $0.01$ ($\nu_s=5$ GHz) up to $0.2$ ($\nu_s=50$ GHz); above $\nu_{s}\gsim30$ GHz the IGM contribution dominates over ISS modulation. We compare our model with data from a $0.3\leq z\leq 2$ quasar sample observed at $\nu_{\obs}=8.4$ GHz. For this high frequency ($10.92\leq \nu_s \leq 25.2$), high galactic latitude sample ISS is negligible, and IGM scintillation can reproduce the observed modulation with a $4\%$ accuracy, without invoking intrinsic source variability. We conclude by discussing the possibility of using IGM scintillation as a tool to pinpoint the presence of intervening high-$z$ groups/clusters along the line of sight, thus making it a probe suitably complementing Sunyaev-Zeldovich data recently obtained by~\textit{Planck}.
\end{abstract}


\section{Introduction}
Scintillation is an optical effect arising when light rays emitted by a compact source pass through a turbulent ionized medium. As the local value of the refraction index varies altering the wave phase, this can lead to geometrical or physical optics effects, such as image distortion or displacement, formation of multiple images, or a variation of the radiation intensity. 
This physical process has been extensively studied for optical radiation seen by a ground telescope through the atmosphere, and radio waves, i.e. the so-called Inter Stellar Scintillation (ISS), in which a pulsar is observed through the turbulent Inter Stellar Medium (ISM) of the Galaxy. These studies have allowed to characterize the spatial distribution of the Galactic ionized medium on a wide range of scales~\citep[e.g.][]{Rickett:2006}; in addition, they have shown that the power spectrum of density irregularities in such diffuse component  has a power-law dependence on the wavenumber which resembles the one expected in the inertial range for a fully developed Kolmogorov turbulence~\citep{Armstrong:1981}.

The intergalactic medium (IGM) is another cosmic component to which scintillation theory could be successfully applied to complement quasar absorption-line experiments probing the neutral component~\citep{Rauch:1998}. However, it is now clear that most of the baryons in the low-redshift Universe reside in an elusive warm/hot ionized medium (WHIM): scintillation, together with X-ray observations, might then represent the optimal tool to infer the WHIM properties.

Additional motivations come from the recent theoretical and observational efforts that firmly established that the IGM is in a turbulent state. From the observational point of view, a direct way to measure the turbulence in the IGM is to look for velocity differences on the smallest spatial scales accessible to observations. To this aim~\citet{Rauch:2001} in a pioneering experiment used lensed quasars in order to observe adjacent C IV profiles in paired lines of sight. According to their findings, velocity differences of $\approx 5$ kms$^{-1}$ on scales of 300 pc have been measured at redshift $z \approx 2.7$. Intriguingly, the inferred turbulent dissipation time-scale is compatible with that of turbulence injection by supernova feedback. This mechanism was investigated, among others, by \citet{Fangano:2007} who simulated a star-bursting analogue of a Lyman-break galaxy at $z\sim 3$ to derive the absorption signatures of the wind environment.   
In~\citet{Evoli:2011} we derived the turbulent Doppler parameter related to galactic feedback 
in the absorbers associated to progenitors of a $10^{13}$~M$_\odot$ galaxy at redshift $z=0$. According to that study, the mean turbulent Doppler parameter peaks at $z \sim 1$ at about $1.5$~km~s$^{-1}$ with a filling factor of $10-20$\% and it slightly decreases at higher redshifts. 

Another hint of a significant turbulent contribution to the IGM kinetic budget comes from the absorption features of different elements. By comparing cosmological simulations with \HI and \HeII Ly$\alpha$ transmitted flux measured in the HE 2347-4342 QSO spectra, \citet{Liu:2006} found that turbulent broadening provides the dominant contribution to the Doppler parameters in the redshift range $2 < z < 2.9$. From a detailed analysis of \OVI and \CIV systems at $z = 2.3$, detected in the VLT/UVES spectra of 18 bright QSO,~\citet{Muzahid:2012} determined a non-thermal broadening in the range $3.6-21.2$~km~s$^{-1}$, with a median value of $8.2$~km~s$^{-1}$, a factor $\sim$2 higher than observed at higher redshifts. In general, the median Doppler parameters measured in the Ly$\alpha$ forest are significantly larger than those predicted by cosmological simulations~\citep{Oppenheimer:2009}. Again, this implies that some energy in non-thermal form must be injected in the gas to explain the observed line broadening~\citep{Meiksin:2001}.
Turbulence can also be produced by the vorticity cascade originating at cosmological shocks, as suggested via cosmological simulations by~\citet{Ryu:2008} and~\citet{Zhu:2010} who derived the average magnetic field strength and turbulent pressure in overdense IGM regions outside clusters/groups. More recently, \citet{Iapichino:2011}~showed that turbulence can be sustained in the intracluster medium by merger-induced shear flows, and by shock interactions in the WHIM.

The idea of using quasar refractive scintillation to infer the properties of the ionized intergalactic/intracluster gas has been proposed by~\citet{Hall:1979} and~\citet{Ferrara:2001}. These authors found that a cluster located at $z < 0.02$ would produce a source r.m.s. intensity fluctuation at $50-100$~GHz, i.e. larger than the galactic signal. If confirmed, scintillation would then represent a method complementary to the standard X-ray emission and Sunyaev-Zel'dovich effect approaches to study the properties of the turbulent intracluster medium. In spite of the potential for IGM studies, intergalactic scintillation has received very little attention. A possible reason is that the IGM scattering measure of a source located at redshift $z$ through a uniformly distributed IGM amounts to $2.8 \times 10^{-5} (1+z)^{9/2}$ times the galactic contribution, and it is detectable in practice only for very distant quasars. However, the cosmic density shows large density fluctuation arising from gravitational instability (the ``cosmic web'') which greatly boost the previous estimate. 
Our aim here is to calculate in detail, using high-resolution, adaptive mesh refinement cosmological\footnote{Throughout this paper we use a WMAP7 cosmology~\citep{Larson:2011} with $\Omega_{\Lambda}= 0.727$, $\Omega_{dm}= 0.228$, $\Omega_{b}= 0.045$, $n=0.967$, $\sigma_{8}=0.811$, $h = 0.704$.} hydrodynamical simulations, the IGM scintillation patterns imprinted in the light received from high-$z$ quasars accounting for the concentrations of hot gas arising in the late phases of cosmic structure formation.

The paper is structured as follows. In Section \ref{ref_sezione_thinscreen} we briefly review the theory of scintillation and, in particular, we discuss the single screen approximation used to model the ISS scintillation. Then in Section \ref{ref_sezione_teoriaIGM} we extend the scintillation model to a continuous medium to which the screen approximation cannot be applied. Section \ref{ref_ISM_scintillation} is dedicated to validate the model against ISS observations. Section \ref{ref_sezione_cosmo} presents the cosmological simulations that we use as an input to calculate the IGM scintillation. Section \ref{ref_sezione_analisi} is devoted to the analysis of the results. In Section \ref{ref_sezione_comparison} we compare our findings with recent observations, and, in Section \ref{ref_sezione_conclusioni}, we state our conclusions.
\section{Scintillation Theory: basics}\label{ref_sezione_thinscreen}
We describe the theory of scintillation starting from the basic definitions and specializing the problem to the thin screen case. We encourage the reader interested in the general theory to read \citet{Wheelon:2001} and \citet{Wheelon:2003} which cover the Born approximation and its extension, the Rytov approximation, respectively. 

Scintillation can be treated in conceptually different ways depending on the dominant effect, namely refraction or diffraction. The former is based on physical optics and is due to interference among multiple ray paths from the source to the receiver; the principle is similar to interferometry as each path forms a different sub-image to be viewed as a slit on an imaginary plane perpendicular to the unperturbed line of sight (l.o.s.) crossing the screen. Diffractive scintillation is a geometrical optics effect due to focussing/defocussing of light, which leads to a random magnification of sub-images of the source. These effects have different time scales and different frequency ranges; we concentrate on refractive scintillation since we will show later on that diffractive effects are negligible for the case of interest here.
\subsection{Formalism}\label{ref_formalism}
Let us consider a light source and an observer connected by a l.o.s. passing through an ionized medium, considered as an ideal plasma of spatial extent $l$. The electromagnetic wave is governed by the Maxwell equations. Away from the source, following \citet{Wheelon:2001}, such equations can be reduced to a single optical equation relating the electric field $\mathbf{E}$ and the current $\mathbf{J}$ in Fourier space:
\begin{equation}\label{eq_ottica}
	\left(\nabla^{2}+\kappa^{2}\epsilon  \right)\mathbf{E}=-4\pi i \kappa \mathbf{J}\, ,
\end{equation}
where $\kappa$ is the wavenumber and $\epsilon$ the dielectric constant. The optical equation (eq. \ref{eq_ottica}) can be studied in the Born approximation describing the electric field with a phase (the iconal, $\Psi$) and an amplitude, $\mathbf{A}$, such that
\begin{equation}
	\mathbf{E}=\mathbf{A}\exp\left( i\kappa \Psi\right)\, .
\end{equation}
Expressing $\mathbf{A}$ as a power series of $\kappa$ in eq. \ref{eq_ottica} and grouping together terms with equal power, one can solve the optical equation order by order. According to the Born approximation \citep{Wheelon:2001} we only retain the leading term in $\kappa$, yielding
\begin{equation}
	\left(\mathbf{\nabla}\Psi\right)^{2}=\epsilon\, .
\end{equation}
Thus the phase can be written as an integral along the l.o.s.
\begin{equation}
	\phi=\kappa\Psi=\kappa\int n\,\mbox{d}s\, ,
\end{equation}
where $n=\epsilon^{1\slash 2}$ is the refraction index. For an ideal plasma\footnote{Here we are implicitly neglecting the induced magnetic field.} the dispersion relation is
\begin{equation}
n^{2}(t,\mathbf{\kappa})=1-\left(\frac{\omega_p}{\omega}\right)^{2} \, ,
\end{equation}
which involves the plasma frequency $\omega_{p}\equiv\sqrt{n_{e}e^{2}\slash m_{e}\epsilon_{0}}$, where $n_{e}$, $m_{e}$ and $e$ are respectively the electron number density, mass and charge, while $\epsilon_{0}$ is the vacuum dielectric constant. The stochastic nature of the problem arises from the link between the refraction index and the underlying turbulence through the electron density.

Assuming the ergodic theorem, turbulence is properly accounted for by the so-called structure phase function, 
\begin{equation}\label{definizioneDphi}
D_{\phi}(\mathbf{r}_{1}-\mathbf{r}_{2})=
\left\langle\left[\phi(\mathbf{r}_{1})-\phi(\mathbf{r}_{2})\right]^{2}\right\rangle \, .
\end{equation}
$D_{\phi}$ represents the phase difference perceived by adjacent observers and averaged over a finite sampling length. Eq. \ref{definizioneDphi} must be expressed in terms of $\Phi_{N}$, the density power spectrum of the medium; one can exploit the finite scale range of turbulence by separating $\Phi_{N}$ into a large scale, time-independent term, $C_{N}(s)$, and a small-scale turbulent term,  $P(\mathbf{k})$, such that
\begin{equation}
\Phi_{N}(\mathbf{r})=C_{N}^{2}(\mathbf{l}) P(\mathbf{k}) \, .
\end{equation}
$C_{N}(\mathbf{l})$ is conveniently defined in terms of the scattering measure, SM, along the l.o.s. normalized to the nominal galactic value $\mbox{SM}_{g}$:

\begin{equation}
	\mbox{SM}_{-3.5}=\frac{\rm{SM}}{\rm{SM_{g}}}=\frac{\int C_{N}^{2}(\mathbf{l})\,\mbox{d}s}{10^{-3.5} \mbox{ m}^{-20/3} \mbox{ kpc}} \, ,
\end{equation}
which can then be written as a function of $n_{e}$ as \citep[e.g.][]{Goodman:1997}:
\begin{equation}\label{eq_sm_goodman}
	\mbox{SM}_{-3.5}=\int\left[\frac{n_{e}(\mathbf{l})}{0.02\mbox{ cm}^{-3}}\right]^{2} \frac{\mbox{d}s}{\kpc}\, .
\end{equation}
For monochromatic plane waves of wavelength $\lambda=c\omega^{-1}=2\pi c\nu^{-1}$ and an isotropic power spectrum (i.e.~$P=P(k)$), $D_{\phi}$ depends only upon $r=\left|\mathbf{r}_{1}-\mathbf{r}_{2}\right|$ and can be written as:
\begin{equation}\label{d_phi2}
	D_{\phi}(r)=\pi^{2}r_{e}^{2} \mbox{SM}\int\left[1-J_{0}(k r)\right]P(k) k\,\mbox{d}k \, ,
\end{equation}
where $J_{0}$ is the $0$-th order Bessel function and $r_{e}$ the classical electron radius.
The phase structure function gives direct information on the spatial\footnote{By the ergodic theorem this is equivalent to the temporal variation.} variation of the intensity, $I$, in terms of the electric field $E$ when adopting the Born approximation~\citep{Wheelon:2001}: 
\begin{equation}
	I\left(r \right)=\langle \mathbf{E}^{*}\mathbf{E}\rangle= E_{0}^{2}\exp\left(-\frac{D_{\phi}\left(r \right)}{2}\right),
\end{equation}
where the subscript $0$ indicates the unperturbed field. An immediate result is that light received at points separated by a distance $r$ is mutually coherent only if $D_{\phi}(r)<1$; thus it is natural to define the field coherence length $s_{d}$ as:
\begin{equation}\label{def_coherence_length}
D_{\phi}\left(s_{d}\right)=1,
\end{equation}
in terms of which we define the diffraction angle (the analogous of the atmospheric seeing in optics) $\theta_{d}\equiv s_{d}\lambda^{-1}$.  Another directly observable quantity is the modulation index defined as the r.m.s. of the intensity autocorrelation:
\begin{equation}\label{intensitycovariance}
m_r \equiv \sqrt{\left\langle\frac{ I\left(r\right)^{2}-\left
\langle I\left(r\right)\right\rangle^{2}}{\left\langle I\left(r\right)\right\rangle^{2}}\right\rangle} \, .
\end{equation}
Note that the Rytov approximation is required to define this moment of the radiation self-consistently, as in the Born approximation the logarithmic intensity variation of the electric field must be vanishing small by definition \citep{Wheelon:2003}.
\subsection{Turbulence power spectrum}\label{app_turbolenza}
We adopt a form of the power spectrum consistent with the interpretation of the ISM pulsar scintillation observations \citep{Rickett:1990}:
\begin{equation}\label{kolgomorovovunque}
P(k)=\left(k^{2}+k^{2}_{out}\right)^{-{\beta}/{2}}\exp\left(-\frac{k}{k_{in}}\right)\, ,
\end{equation}
where $\beta$ is a free parameter, $k_{in}^{-1} \approx 10^{12}$ cm ($k_{out}^{-1} \approx 10^{18}$ cm) is the inner (outer) scale  
in the ISM~\citep{Lambert:2000}. In the inertial range eq. \ref{kolgomorovovunque} reduces to
\begin{equation}\label{kolgomorovbeta}
P(k)=k^{-\beta}\quad\mbox{for}\quad k_{out}\ll k\ll k_{in}
\end{equation}
Using eqs. \ref{d_phi2} and \ref{kolgomorovbeta}, we can find an analytical expression for $D_{\phi}$ for the plane wave case
\begin{equation}
D_{\phi}(r)=\pi^{2}r_{e}^{2}f(\beta) r^{\beta-2} \mbox{SM },
\end{equation}
where the dimensionless factor $f(\beta)$ can be expressed in terms of the $\Gamma$ function,
\begin{equation*}
f(\beta)=\left\{
\begin{aligned}
 \frac{8\,\Gamma\left(\beta\slash2\right)\Gamma\left(2-\beta\slash2\right)}{(\beta-2)2^{\beta-2}}\quad&\mbox{ } r>k_{in}^{-1}\\
\Gamma\left(2-\beta\slash2\right)k_{in}^{-\beta}\quad&\mbox{ }  r <k_{in}^{-1}\, .
\end{aligned}
\right.
\end{equation*}
By using the $\Gamma$ function properties we further get:
\begin{equation}
D_{\phi}(r)=8\,\pi^{3}r_{e}^{2} \frac{1-\beta\slash2}{\beta-2}\frac{2^{2-\beta}}{\sin\left(\pi\frac{\beta}{2}\right)} r^{\beta-2}\mbox{SM}\, ,
\end{equation}
in the inertial range. In particular, assuming a Kolmogorov power spectrum for the turbulence (namely $\beta={11}/{3}$), and using the definition given in eq. \ref{def_coherence_length}, we can write the characteristic diffraction angle as
\begin{equation}\label{angolodiff}
\theta_{d}=2.93 \, \nu_{10}^{-{11}/{5}} \mbox{SM}_{-3.5}^{{3}/{5}} \,\, \mu\mbox{as} \, ,
\end{equation}
where $\nu_{10} \equiv \nu / 10\mbox{ GHz}$.
Lacking precise data for the IGM, we adopt for the turbulence spectrum in the IGM the same properties as in the ISM. While this assumption is robust for the spectral index $\beta$ as most astrophysical fluids develop a self-similar Kolmogorov spectrum in their inertial range, its validity is less clear for what concerns $k_{in}^{-1}$ and  $k_{out}^{-1}$ which depend on the energy injection and dissipation mechanisms \citep{Rickett:1990,Goodman:1997}. However, as noted by \citet{Coles:1987}, scintillation is relatively independent from $k_{\rm in}$; nevertheless,  the final result could in principle depend on the choice of $k_{\rm out}$. In fact, any change in $k_{out}^{-1}$ can be seen as a different SM normalization, which we have implicitly absorbed in $\mbox{SM}_{-3.5}$ via eq. \ref{eq_sm_goodman}.

In order to evaluate the systematic uncertainty associated with our choice of $k_{\rm out}$, we have numerically integrated eq. \ref{d_phi2} for the power spectrum given eq. \ref{kolgomorovovunque} and taking the two extreme values $k_{out}^{-1}=\{1, \,100\}\mbox{ pc}$. These values are typically assumed in theoretical modeling of ISS \citep[e.g.][]{Coles:1987} and directly observed in our Galaxy via Faraday rotation measurements~\citep{Minter:1996ApJ}. As a result, we find that our estimates of $\theta_{d}$ through the equation \ref{angolodiff} change by a factor of $\sim5$.

The effect on the final results is even less severe in the weak refractive regime, since the relevant quantities are weakly dependent on the value of $\theta_{d}$ (as we describe in Section \ref{ref_subsec_thin_srceen}). We conclude that, even if IGM turbulence is injected on a significantly different scale with respect to that of the ISM, this will only reflect on our results as a small variation in the overall SM normalization.
\subsection{Thin screen approximation}\label{ref_subsec_thin_srceen}
\begin{figure}
\begin{center}
\ifpdf
  \includegraphics[width=8.3cm,height=4cm]{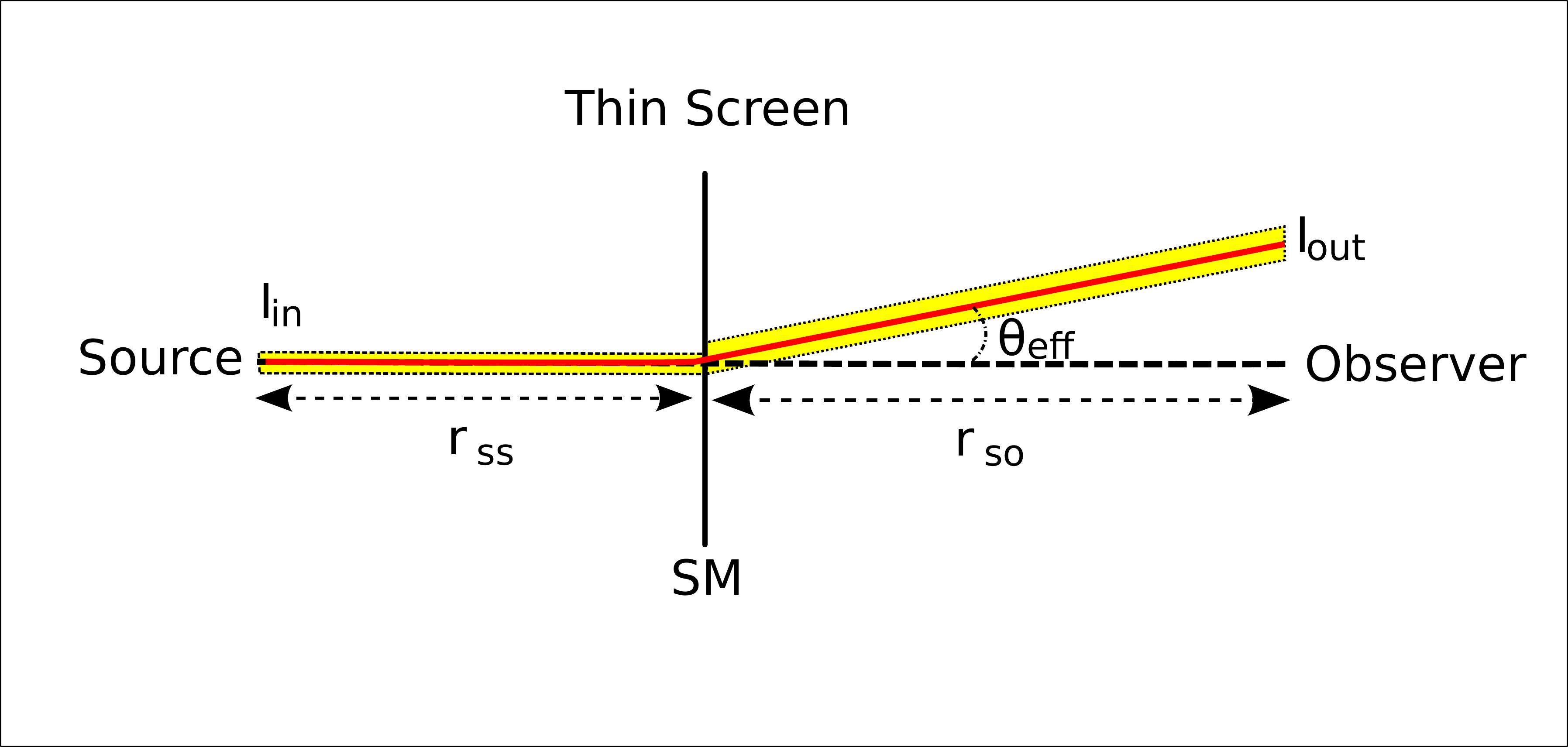}
\else
  \includegraphics[width=8.3cm,height=4cm]{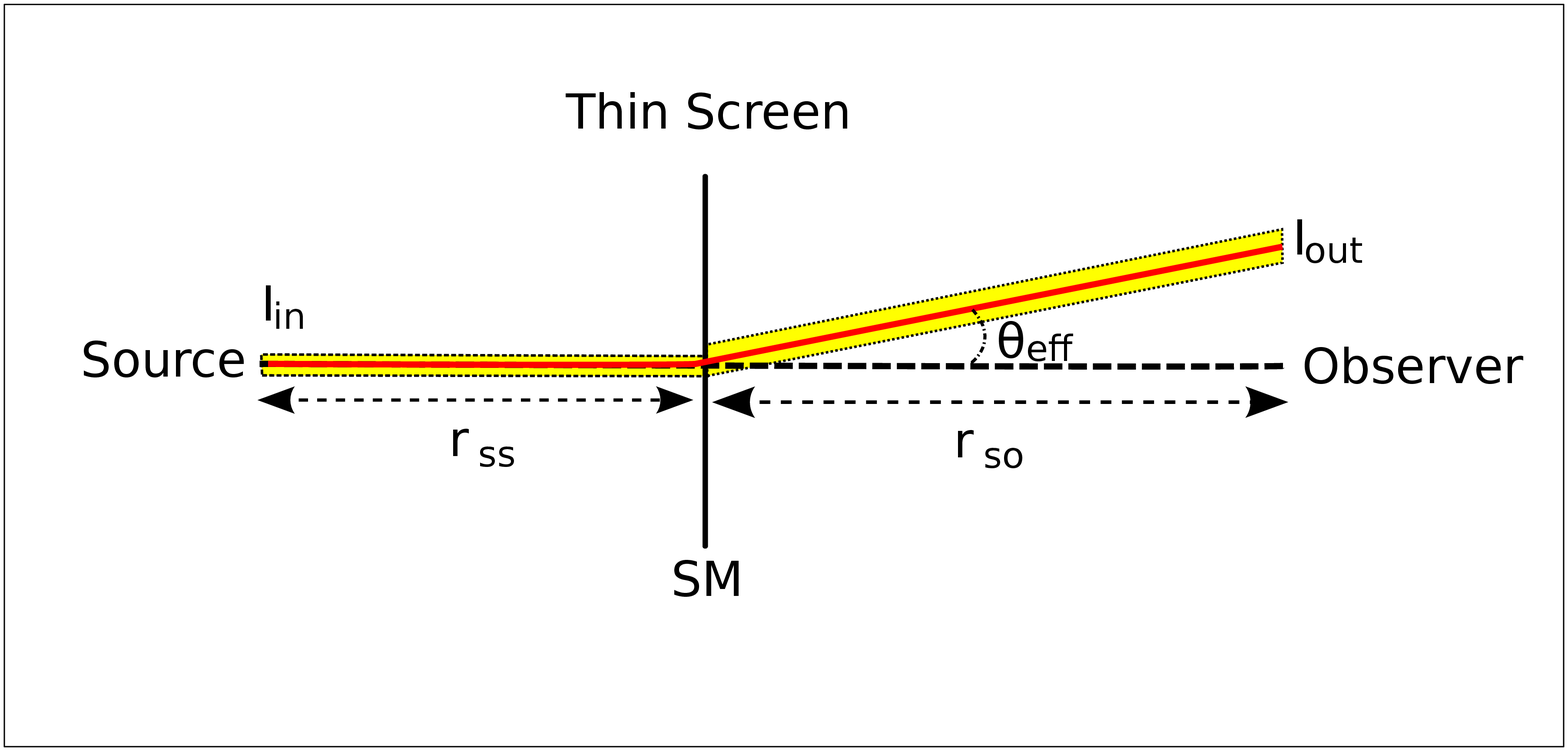}
\fi
\end{center}
\caption{Sketch of the scintillation process from a thin screen: the thick dashed line represents the unperturbed path, while the red line is the actual path given by the screen scattering measure $\mbox{SM}$ and geometrical configuration of the distances $r_{ss}$, $r_{so}$. The resulting angular displacement $\theta_{\eff}$ is given in eq. \ref{thetaefficace}. The intensity modulation  (eq. \ref{m_r-goodman}) is visualized as a variation of the beam thickness (yellow). See \ref{ref_subsec_thin_srceen} for the detailed definitions.
\label{disegnino_sin}}
\end{figure}

When the spatial extent of the scattering medium (the ``screen'') is much smaller than the screen-observer, $r_{so}$, and the source-screen, $r_{ss}$, distances, the problem can be studied using the thin screen approximation, as it is canonically done to study pulsar scintillation through the ISM. For a visual representation of the problem see Fig.~\ref{disegnino_sin}.

As the medium can be considered to be compressed into a plane perpendicular to the l.o.s., refractive effects can be inferred from Fresnel theory, which sets a characteristic radius for the change in the optical properties:
\begin{equation}
r_{F}=\sqrt{\lambda D_{F}}\, ,
\end{equation}
where $D_{F}$ is given by \citet{Lee:1977}
\begin{equation}
D_{F}=\frac{r_{ss}r_{so}}{r_{ss}+r_{so}}.
\end{equation}
By introducing a modified distance $\tilde{d}=r_{so}^{2}/D_{F}$ one can write the Fresnel angle as
\begin{equation}\label{eq_thetaf}
\theta_{F}\simeq2.57 \, \nu_{10}^{-{1}/{2}} \tilde{d}_{\kpc}^{-{1}/{2}}\, \mu\mbox{as}\, .
\end{equation}
Together with $\theta_{d}$, the Fresnel angle can be used to compute the effective angle, $\theta_{\eff}$, which can be understood as the point spread function of a source of angular size $\theta_{s}$ seen through the scattering medium \citep[i.e.][]{Goodman:1997}:
\begin{equation}\label{thetaefficace}
\theta_{\eff}=\sqrt{\theta_{s}^{2}+\left(0.71\,\theta_{d}\right)^{2}+\left(0.85\,\theta_{F}\right)^{2}}\, ,
\end{equation}
and the modulation index can then be written as
\begin{equation}\label{m_r-goodman}
m_{r}=0.114 \,\nu_{10}^{-2}\tilde{d}_{\kpc}^{-{1}/{6}}\left(\frac{\theta_{\eff}}{10\ \mu\mbox{as}}\right)^{-{7}/{6}}\mbox{SM}_{-3.5}^{{1}/{2}}\, .
\end{equation}
Notice that, using the same argument given in Sec. \ref{app_turbolenza} for $\theta_d$, the variation $k^{-1}_{out}=\{1,100\}$~pc would lead to a change of $\sim3$ for $m_r$, under the condition $\theta_d\gg\theta_{F},\theta_{s}$.

The refractive scintillation regime is defined by the following inequality: 
\begin{equation}\label{condizionerifrattiva}
\theta_{s}<\theta_{d}\, ;
\end{equation}
for a point source $\theta_{s}\ll 1$, and the resulting effect is a displacement of the image, while for an extended one it corresponds to a distortion. Since refractive scintillation is incoherent, the relevant fluctuations are those with wavelength larger than the projection on the scattering screen, or $\lambda > \theta_{\eff}D_{F}$; the characteristic refractive time scale can then be obtained from 
\begin{equation}\label{trifdef}
t_{\refr}=\frac{D_{F}}{v_{\perp}}\theta_{\eff},
\end{equation}
using $v_{\perp} \approx c_s$, i.e. the sound speed in the medium \citep{Ferrara:2001}.
\section{Beyond the thin screen limit}\label{ref_sezione_teoriaIGM}
The ISS treatment based on the thin screen approximation cannot be directly applied to a high-redshift quasar scintillating through the ionized IGM. In this case the two following assumptions are in general no longer valid: (i) the physical properties of the scattering medium must be slowly varying along the l.o.s.; (ii) the spatial extent of the medium must be negligible with respect to $r_{ss}$ and $r_{so}$. This conditions do not hold in the typical situation in which the l.o.s. intersects a number of large density fluctuations separated by voids. 
In fact (see Fig. \ref{densita_fig_3_0}) the density along a typical l.o.s. of our simulation can vary by more than one order of magnitude on scales of $\sim 10$~Mpc~$h^{-1}$. In this case, condition (i) is clearly violated, making difficult to identify a unique scintillation thin screen. As a consequence also condition (ii) cannot be applied in this case.
\begin{figure}
\begin{center}
\ifpdf
  \includegraphics[width=8.3cm,height=4cm]{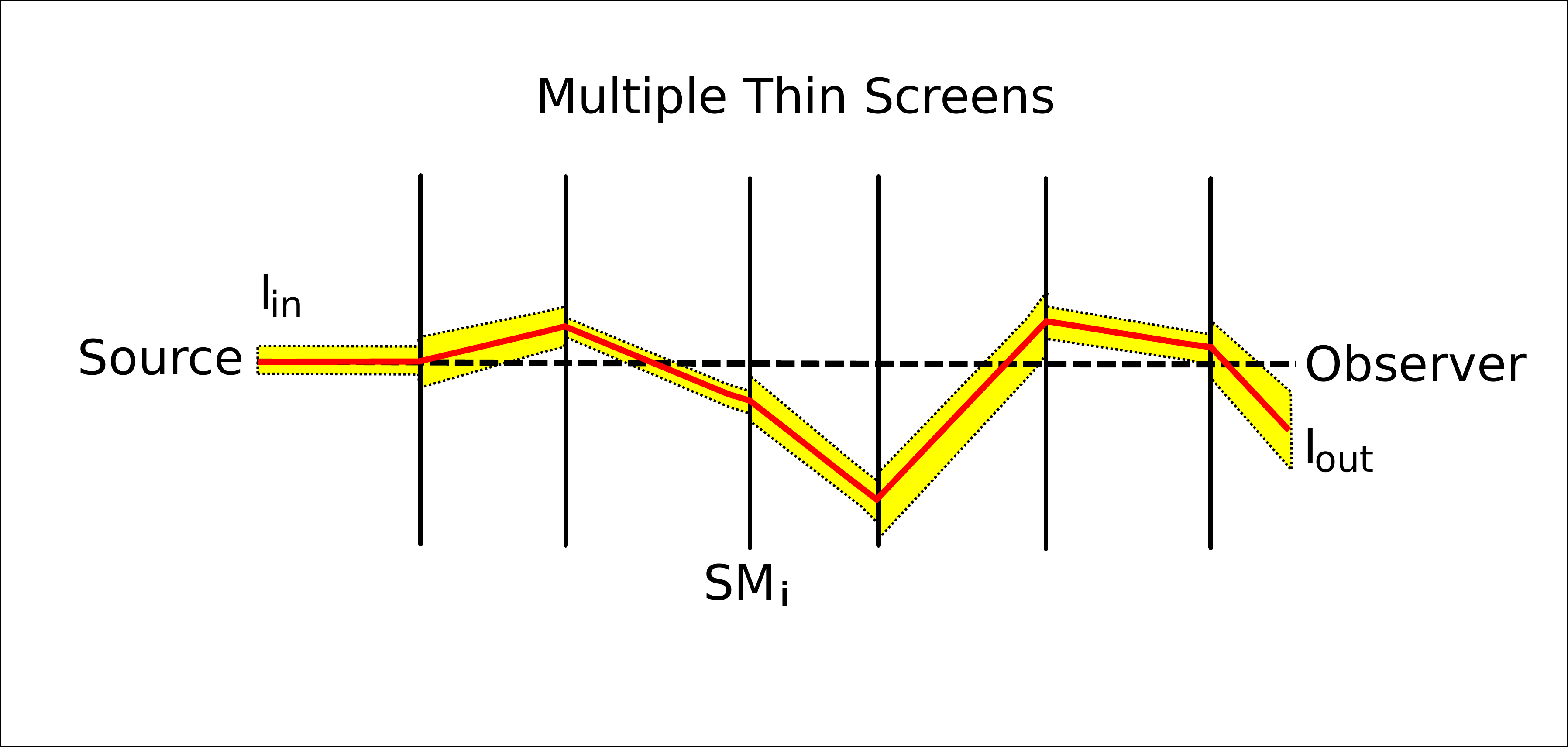}
\else
  \includegraphics[width=8.3cm,height=4cm]{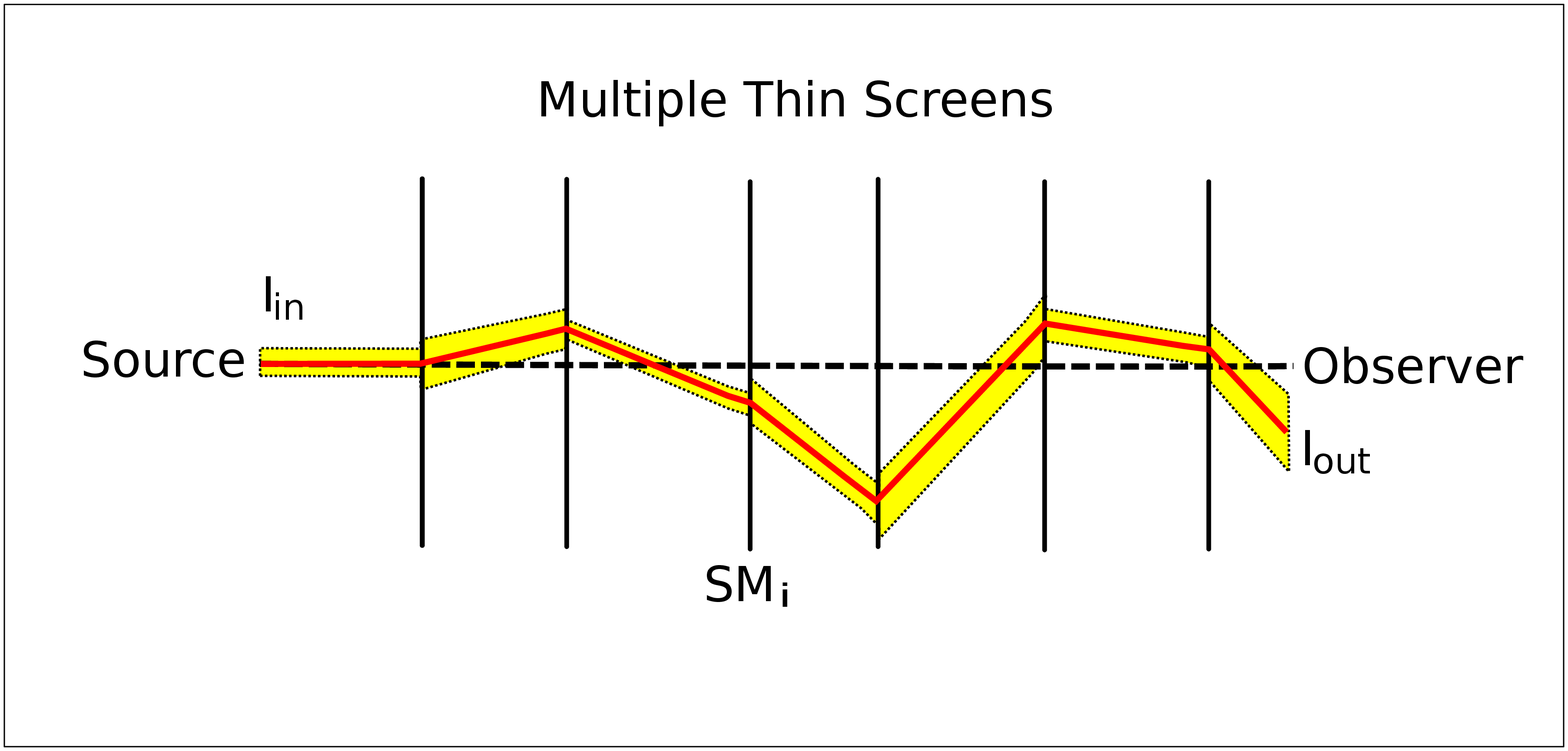}
\fi
\end{center}
\caption{Sketch of the refractive scintillation process through a series of thin screens with scattering measure $\mbox{SM}_i$. Note that at each screen the direction and the intensity of the light ray is modified.
	\label{disegnino_mul}}
\end{figure}

Note that the thin screen limit is based on the optical equation (eq. \ref{eq_ottica}) solution with the Born approximation\footnote{This argument holds also for the Rytov approximation; for details see \citep{Wheelon:2003}.}, thus retaining only first order terms of the amplitude of the electric field. Similarly to a scattering problem, higher orders correspond to progressively multiple scattering (lower probability) events; with increasing path length the contribution of these terms becomes more important in the solution of eq. \ref{eq_ottica}.

To overcome these problems we have devised a strategy in which we split the medium into turbulent layers treated as thin screens, and describe their collective effect through a physically motivated interaction among them. We make sure that this method reproduces the correct behavior in the thin screen regime; in addition it should also provide a good description of the process when the thin screen assumption breaks down.

Before explaining the details of the method, we highlight two important simplifications. First, inspired by ISS, we assume that IGM scintillation is in the refractive regime for $\nu_{10}\gtrsim 5$ \citep{Rickett:1990}. Second, as the IGM scattering measure is typically smaller than the ISM one \citep[i.a.][]{Goodman:1997}, implying small angular deviations from the ray path. Formally, this is equivalent to say that we can safely assume that IGM scintillation can be studied in the forward scattering approximation. 

\subsection{Setting the screens}
We define a thin screen in a continuous medium as a density layer in which statistically at least one refraction event takes place. This might be thought, in analogy with radiative transfer, as the condition for a unity scintillation optical depth. In turn this requires the condition expressed by eq. \ref{condizionerifrattiva}, $\theta_{d} > \theta_{s}$, where $\theta_{s}$ is now the source angular size seen by the screen itself. For a Kolmogorov power spectrum, from eq. \ref{condizionerifrattiva} it follows that an IGM layer of proper size $l$ located at distance $r_{ss}$ from the source can be considered as 
a screen if the following condition is met: 
\begin{subequations}\label{condizionerifrattiva_screens}
\begin{align}
&\theta_{s} < 2.93 \left(\frac{\nu_{s}}{10 \mbox{ GHz}}\right) ^{-{11}/{5}}{\cal I}^{3/5}(r_{ss}, l) \ \mu\mbox{as} \\
&{\cal I}(r_{ss}, l)\equiv\int_{r_{ss}-{l}/{2}}^{r_{ss}+{l}/{2}}\left[\frac{a(s)}{a_{s}}\right]^{{11}/{3}} \frac{\mbox{d}}{\mbox{d}s} \mbox{SM}_{-3.5}\,\mbox{d}s\, ,
\end{align}
\end{subequations}
where $\nu_{s}$ and $a_{s}=1/(1+z_s)$ are the rest frequency and the expansion factor of the source, respectively; additionally we pose that the source angle for all but the first screen is given by the effective diffractive angle produced by the previous screens. Once the thin screens are defined as above, a given light ray along its path to the observer encounters $N$ screens, located at appropriately defined redshifts $z_i$ ($i=1,\dots,N$), each of which can be treated as described in Section \ref{ref_sezione_thinscreen}. The $i$-th screen sees the source through the previous  $(i-1), (i-2),\dots, 1$ screens; the Fresnel radius relative to this screen can then be written recursively:
\begin{equation}
r_{F,i}=\sqrt{\frac{\lambda}{2}\frac{a_i}{a_{s}} \frac{(l_i+l_{i+1})(l_i+l_{i-1})}{l_{i-1}+2l_i+l_{i+1}}}\, .
\end{equation}
Given this definition, one can assign to each $i$-th screen an effective angle $\theta_{\eff,i}$ using eq. \ref{thetaefficace}, which depends on its SM$_i$ and $l_i$ value. 

Additionally we define the \textit{equivalent scattering measure}, $\mbox{SM}_{\equ}$, i.e. the scattering measure obtained by compressing the density on the entire l.o.s. on a thin screen:
\begin{equation}\label{eq_def_smequ}
\mbox{SM}_{\equ}=\sum_{i=1}^{N}\mbox{SM}_{-3.5,i}\, .
\end{equation}

\subsection{Effective screen interaction}\label{ref_subsec_interaction}
To model the effects of a series of thin screens we have implemented an effective screen-screen interaction based on the idea that scintillation can be interpreted as a Levy flight \citep{Boldyrev:2003,Boldyrev:2006}. This corresponds to  a random walk in which the variance of the distribution from which the path increment is drawn is not finite. For the problem of intergalactic scintillation at hand here, this means that for a random l.o.s. to quasar, scintillation is dominated by relatively rare screens located in high density regions (i.e. where structure formation takes place) and hence having a scattering measure far larger than the sum of the others. Ideally the probability distribution for the scattering measure along a l.o.s. should be given by an exponentially truncated Levy distribution, in which the cutoff is determined by cosmic structure formation.
\begin{figure*}
\begin{center}
\ifpdf
  \includegraphics[width=8.3cm]{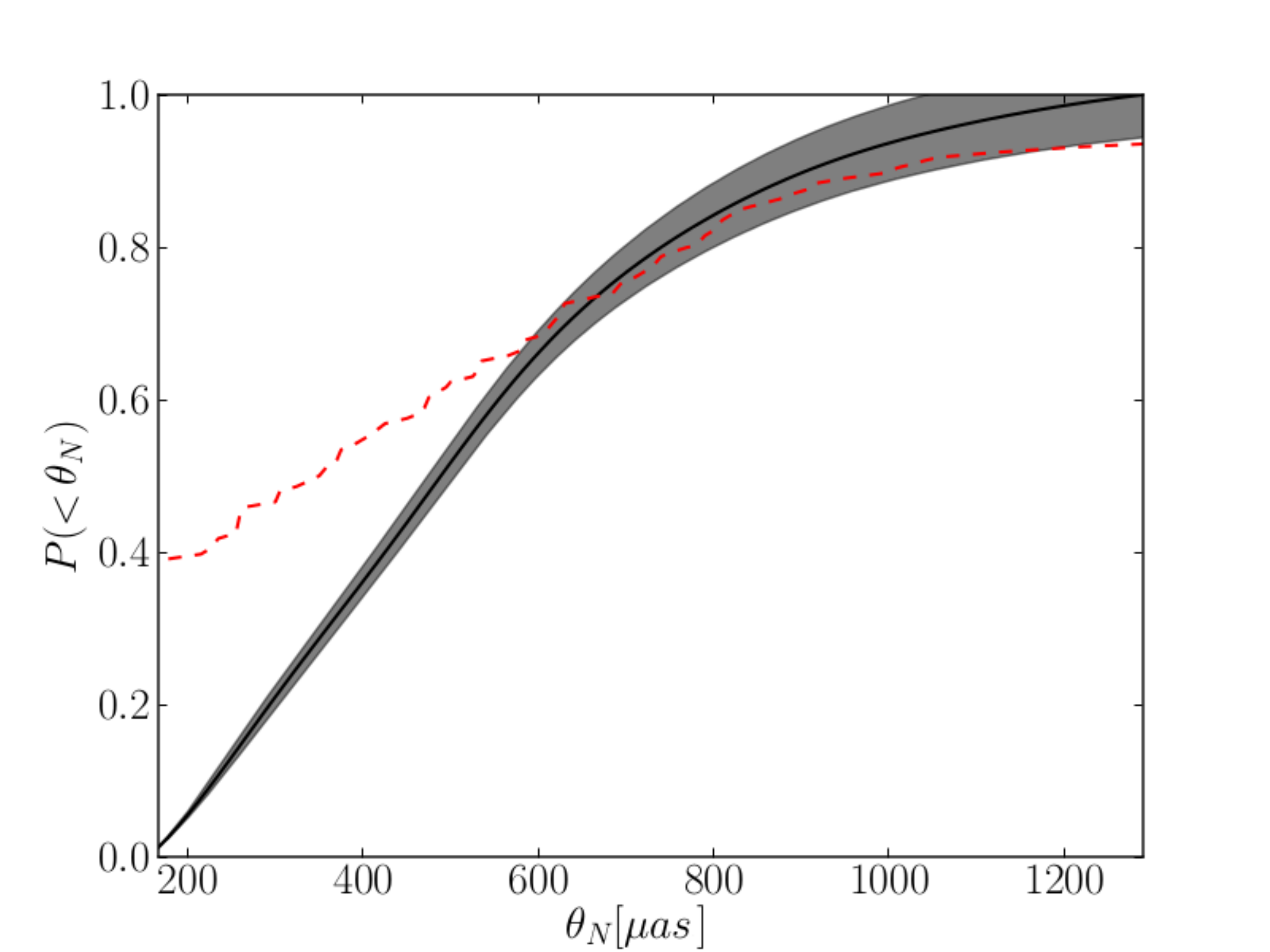}
  \includegraphics[width=8.3cm]{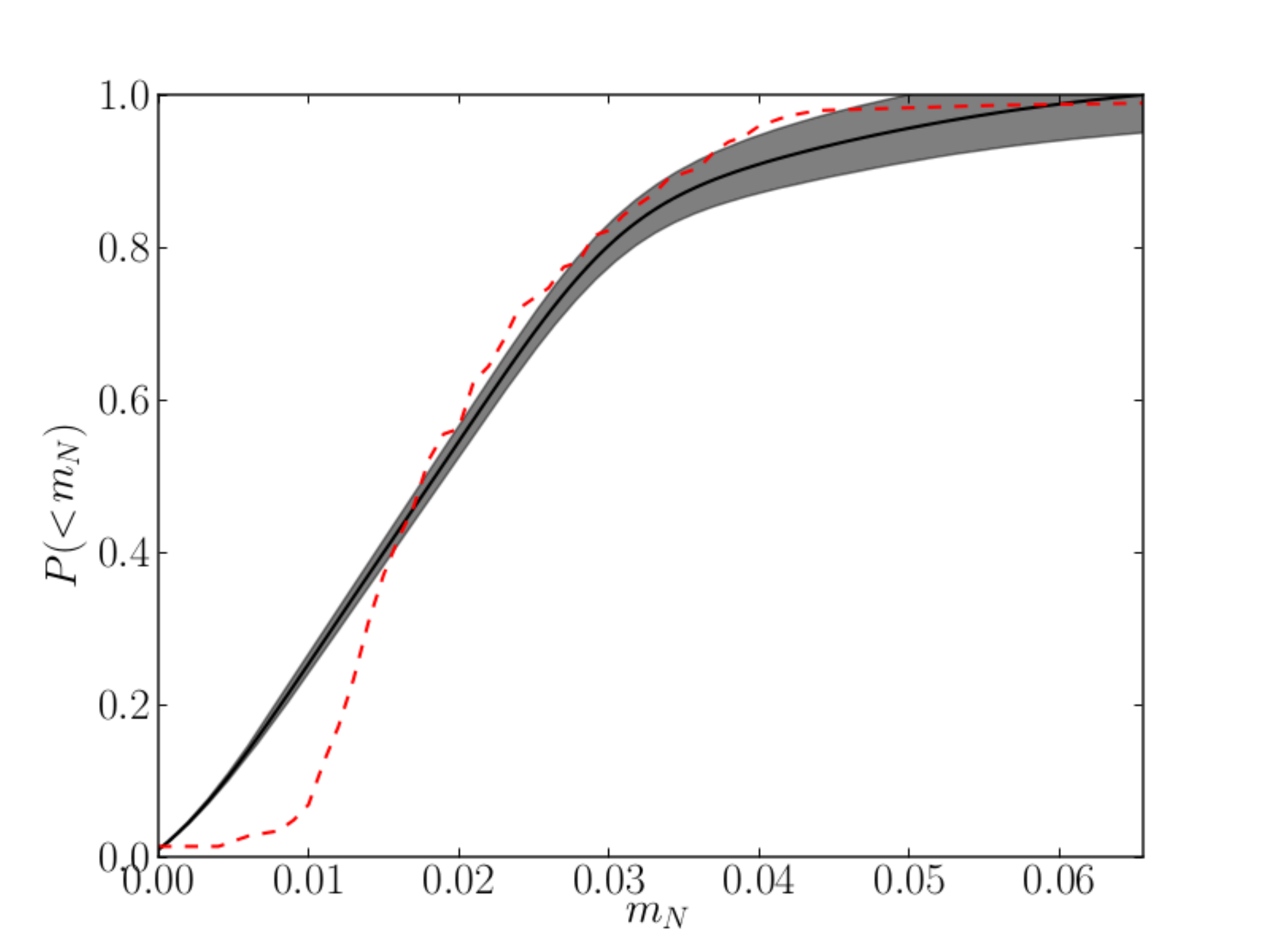}
\else
  \includegraphics[width=8.3cm]{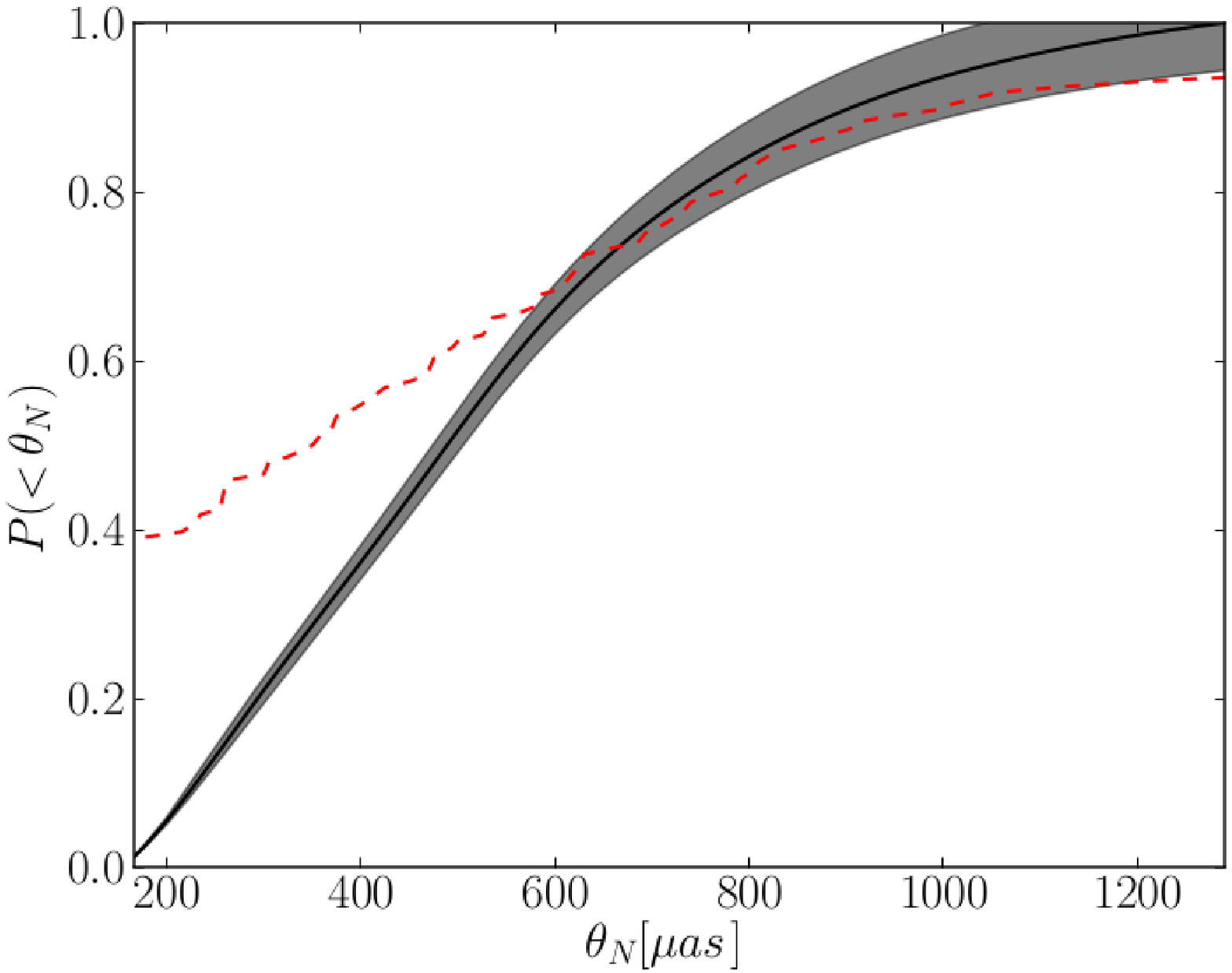}
  \includegraphics[width=8.3cm]{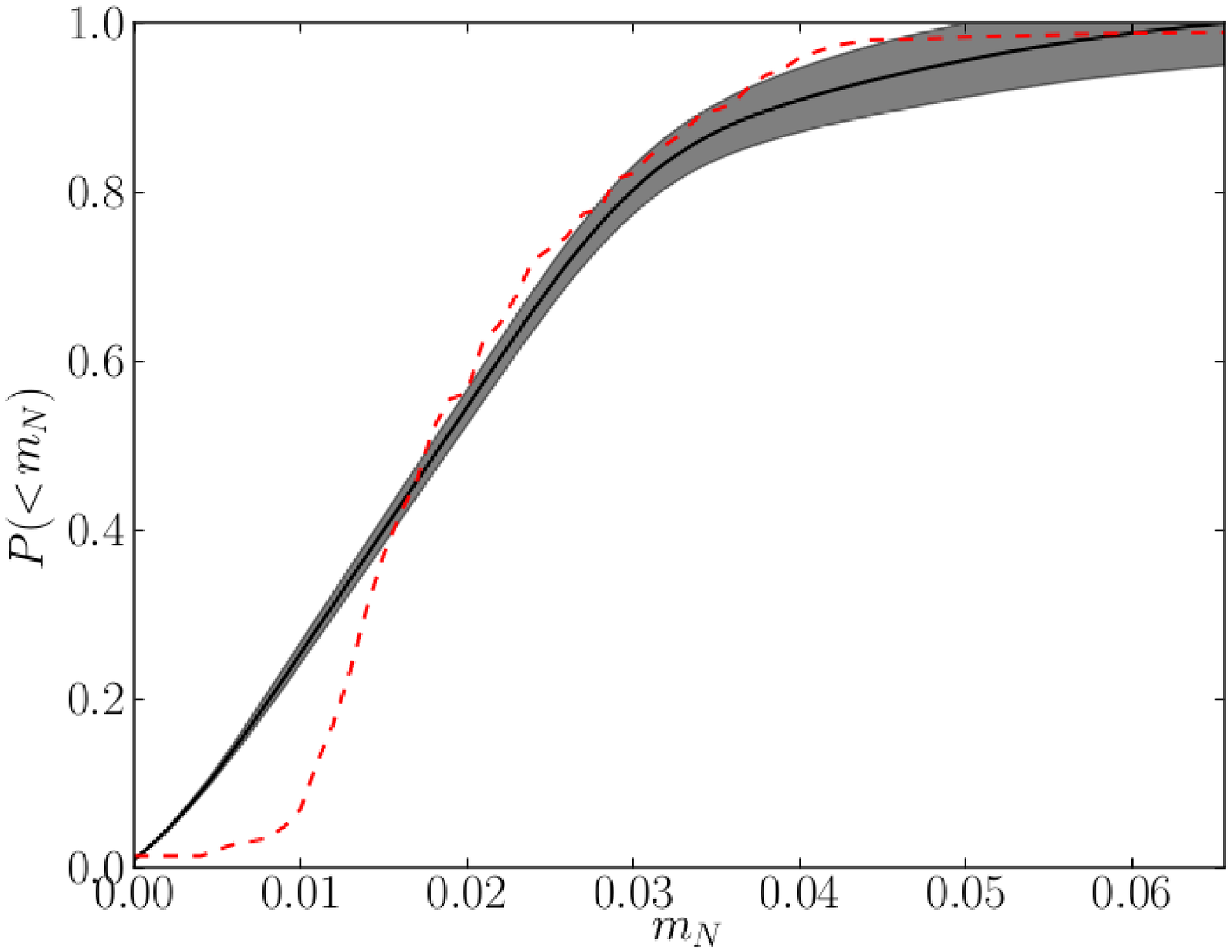}
\fi
\end{center}
\caption{ISM CDF of the refraction angle (left) and modulation index (right) calculated for $T_{\obs}=1$ d, for a source frequency $\nu_{s}=8$ GHz; the shaded region indicates the error of the probability and the red broken line is the CDF inferred from the data in \citet{Rickett:2006}, plotted without propagating the instrumental errors. Details of the calculation are indicated in Appendix \ref{app_PDF}.\label{ISM_mr_theta}}
\end{figure*}

We start by focusing on a single l.o.s.: motivated by the above arguments, we treat the angle $\theta_{\eff,i}$ as the norm of a bidimensional vector laying on the plane perpendicular to the l.o.s., with direction given by a random unit vector
\begin{equation}
\hat{n}_i=(\cos\alpha_i,\sin\alpha_i)\, .
\end{equation}
Since the refraction is the cumulative effect of the screens, the resulting angle at the $i$-th screen is given by the vector sum of the variations produced by the screens $1,2\dots,i$ (see  Fig. \ref{disegnino_mul}). However, as scintillation theory is strictly valid only if the time interval during which the source is monitored, $T_{\obs}$, is longer than the scintillation time scale (see Section \ref{ref_formalism}), the sum must be written as 
\begin{equation}\label{angolorifrazionetotale_1}
\theta_{i}\left(T_{\obs},\alpha\right)=\left|\sum_{j=1}^{i}\theta_{\eff,j}\hat{n}_{j}\chi_j\right|\, ,
\end{equation}
where $\alpha=\left(\alpha_1,\dots,\alpha_N\right)$ is a family of uniformly distributed random parameters, with $\chi_i=\chi\left(T_{\obs}-t_{
efr,i}\right)$ we indicate the step function and the time scale, using eq. \ref{trifdef}, can be written as 
\begin{equation}
t_{
efr,i}=l_i(a_ic_{s,i})^{-1}\theta_{\eff,i}\, ,
\end{equation}
which depends on the sound velocity of the $i$-th screen; although apparently crude, this approximation will be shortly shown to provide a satisfactory description of the process.
Given a l.o.s. and $T_{\obs}$, the actual total refraction angle, $\theta_{N}$, is obtained by averaging over different Monte Carlo realizations of $\alpha$: 
\begin{equation}
\theta_{N}\left(T_{\obs}\right)= \left\langle\theta_{N}\left(T_{\obs},\alpha\right)\right\rangle; 
\end{equation}
the corresponding error is given by 
\begin{equation}
\Delta\theta_{N}\left(T_{\obs}\right)=
\sqrt{\left\langle \left[\theta_{N}\left(T_{\obs},\alpha\right)-  \left\langle\theta_{N}\left(T_{\obs},\alpha\right)\right\rangle \right]^{2} \right\rangle}\, .
\end{equation}
The total modulation index, $m_{N}$, is calculated in two-step process. First we assign a modulation index to the $i$-th screen using the thin screen prescription (eq. \ref{m_r-goodman}):
\begin{equation}
\begin{aligned}
m_{r,i}=&\ 0.114\ \nu_{10}^{-2}\tilde{d}_{i,kpc}^{-1\slash 6}\mbox{SM}_{-3.5,i}^{1\slash 2}\times\\
&\times\left[\frac{ \sqrt{\theta_{\eff,i}^{2}+\theta_{i-1}^{2}(\alpha) }}{10\ \mu\mbox{as}}\right]^{-7\slash 6} . 
\end{aligned}
\end{equation}
Note that $m_{r,i}$ depends on the arrival angle from the $(i-1)$-th screen.

The second step consists in updating the intensity at the $i$-th screen. Indicating with ${\cal G}(\sigma)$ a Gaussian random variable with zero mean and variance $\sigma$, the intensity $I_i$ (see Fig. \ref{disegnino_mul}) can be written as follows
\begin{equation}
	I_i=I_{i-1}\left[1+{\cal G}(m_{r,i})\right]\, .
\end{equation}
Embedding the time dependence in the step function, $\chi_i$, the final intensity for a given realization of $(\alpha,{\cal G})$ can be schematically written as
\begin{equation}
I_{N}\left(T_{\obs},{\cal G},\alpha\right)=I_0\prod_{i=1}^{N}\left[1+{\cal G}\left(\chi_i m_{r,i}\right)\right]\, .
\end{equation}
Finally, according to eq. \ref{intensitycovariance}, the total modulation index is then the average over different $(\alpha,{\cal G})$ realizations 
\small
\begin{equation}\label{m_r_totale}
m_{N}\left(T_{\obs}\right)= \sqrt{\left\langle\frac{ \left(I_{N}\left(T_{\obs},{\cal G},\alpha\right)\right)^{2}-\left
\langle I_{N}\left(T_{\obs},{\cal G},\alpha\right)\right\rangle^{2}}{\left\langle I_{N}\left(T_{\obs},{\cal G},\alpha\right)\right\rangle^{2}}\right\rangle};
\end{equation}
\normalsize
note that $m_{N}$ is independent of $I_0$.

To compute the errors we use a bootstrapping method. First we take a series of $N_{\tot}$ realizations of the random variables $(\alpha,{\cal G})_{1},\dots,(\alpha,{\cal G})_{N_{\tot}}$ out of which we extract $N_{\ext}$ pairs allowing for repetitions. Then we estimate the error by calculating $m_N$ as in eq.~\ref{m_r_totale}, performing the average over the $N_{\ext}$ realizations. In the following we use $N_{\tot}=5\times 10^{3}$ and $N_{\ext}=2 \times 10^{3}$, a choice providing a suitable convergence of the results.

To study the statistical properties of the scintillation on multiple l.o.s. we derive several Probability Distribution Functions (PDFs) and their corresponding moments (for details see Appendix \ref{app_PDF}), typically using $N_{\los}=2\times 10^{4}$ l.o.s. for convergence reasons. 

\section{Model validation}\label{ref_ISM_scintillation}
Before applying the method described so far to the intergalactic scintillation of distant quasars, it is necessary to validate our scheme locally by comparing its predictions with the available experimental data on interstellar scintillation. This is an important step as the model depends on the family of random parameters $(\alpha,{\cal G})$ and the statistical reliability of the results must be
assessed.
\begin{figure}
\begin{center}
\ifpdf
  \includegraphics[width=8.3cm]{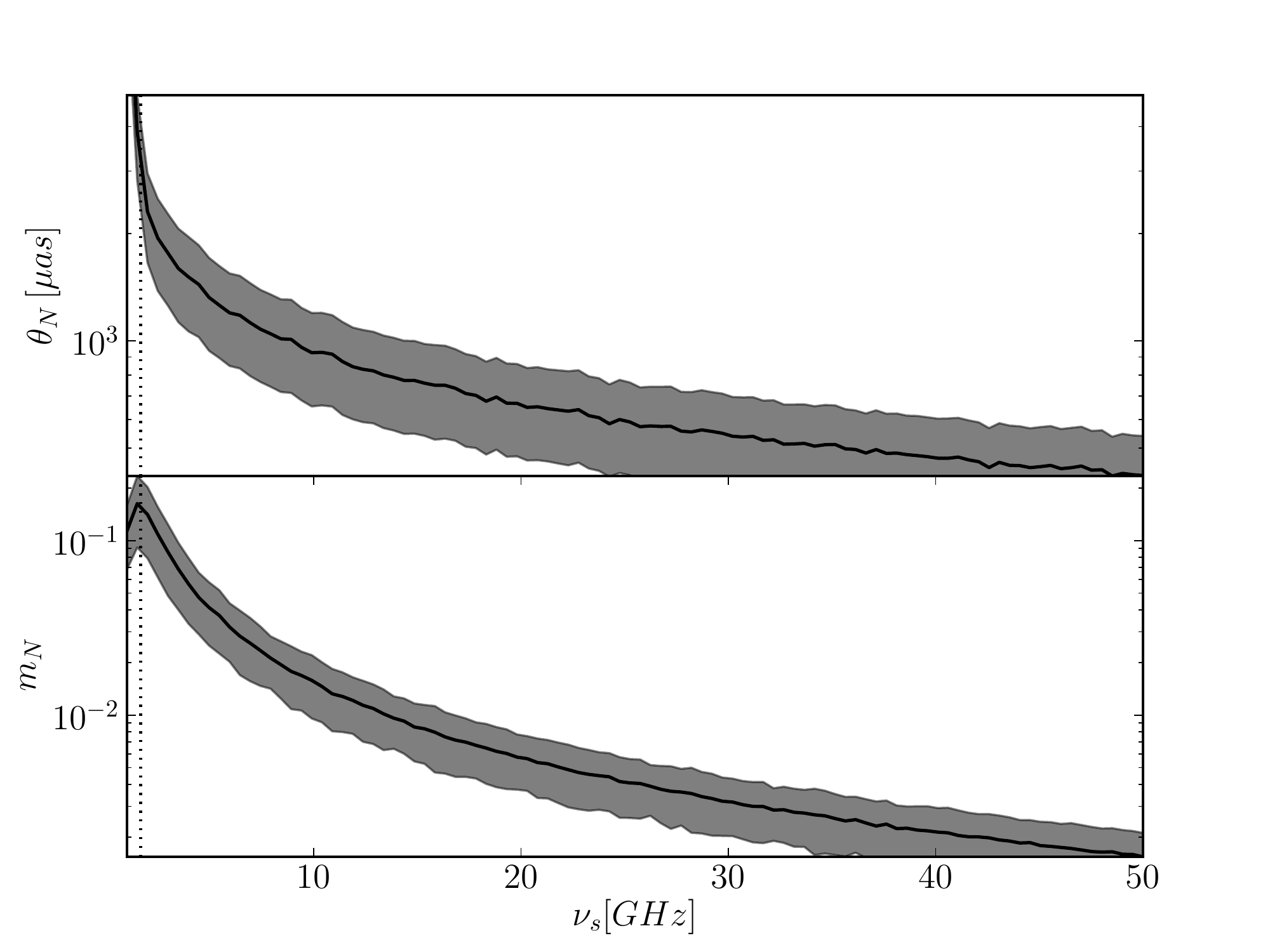}
\else
  \includegraphics[width=8.3cm]{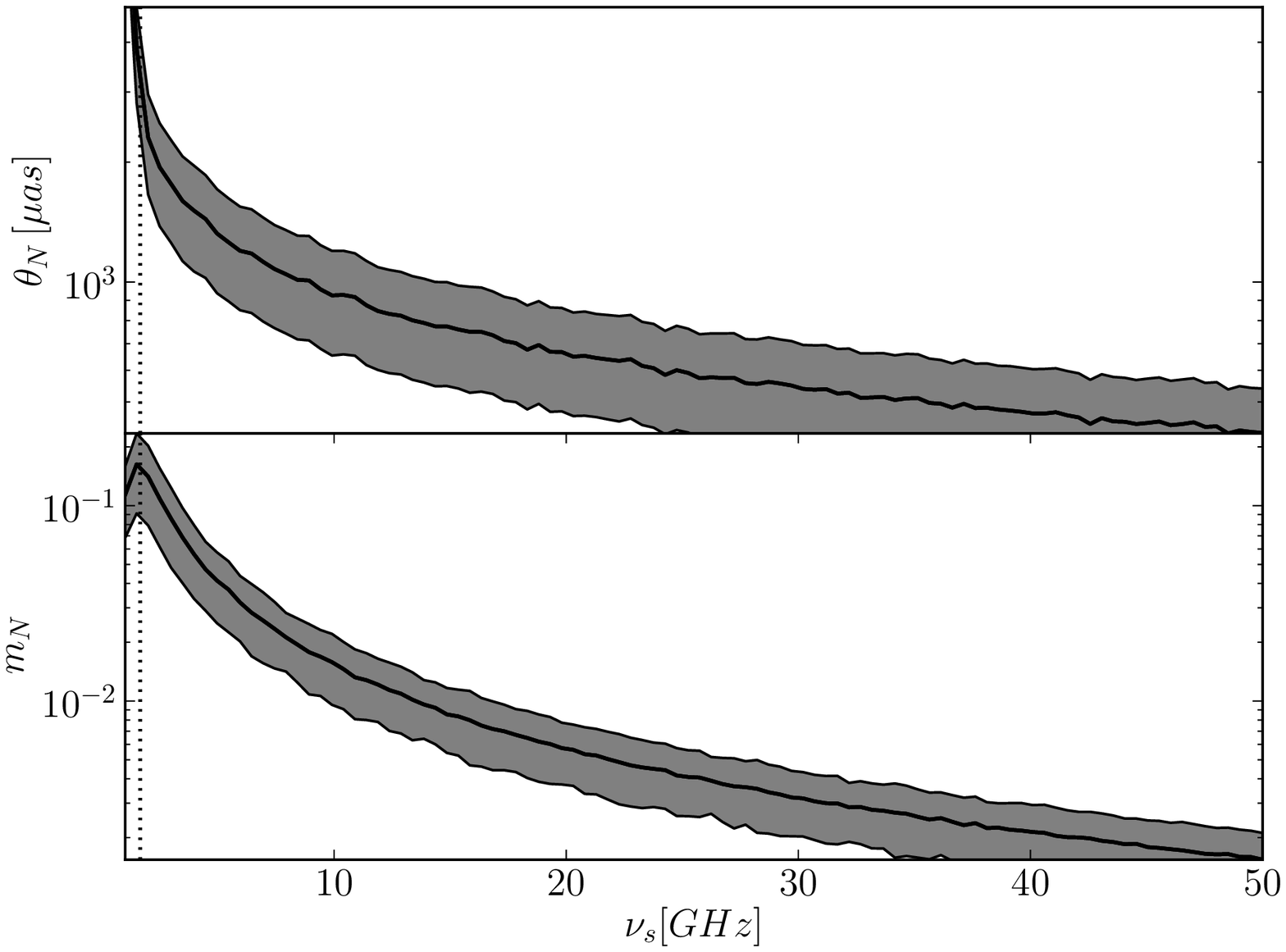}
\fi
\end{center}
\caption{ISM refraction angle (upper panel) and modulation index (lower) as a function of the source frequency $\nu_s$ for lines of sight with $\langle\mbox{SM}_{\equ}\rangle= 0.828$ for $T_{\obs}=1\mbox{ d}$. The solid lines represents the mean while the shaded regions indicate r.m.s. fluctuations. The dotted vertical line marks the critical frequency $\nu_c\simeq2$~GHz (see text).\label{ISM_source_overview_fig}}
\end{figure}

In the Milky Way the observed scintillation is mainly contributed by an ionized layer \citep{Reynolds:1989} whose free electron number density as a function of the height $h_z$ can be approximated by the following analytical expression~\citep{Ferriere:2001}: 
\begin{equation}
{n_{e}(h_z)}=0.15\,e^{-{|h_z|}/{70\rm pc}}+ 0.25\,e^{-{|h_z|}/{900\rm pc}} \, \mbox{ cm}^{-3};
\end{equation}
the \HII layer has a temperature $T\approx 8 \times 10^{3}$K~\citep{Ferriere:2001}. 
We assume that the sources are located on a sphere of radius $r_{so}$=5 kpc centered on the Sun and with Galactic latitude, $b$, uniformly distributed in the interval $[-90^\circ,+90^\circ]$; results are not dependent on the distance choice as long as $r_{so}>5$ kpc, due to the exponential decline of the electron number density. We obtained a mean equivalent scattering measure of $\langle\mbox{SM}_{\rm equ}\rangle = 0.828$.

To validate our scheme we compare our predictions with the \citet{Rickett:2006} data relative to the  ISS of a sample of 146 extra-galactic sources collected from different surveys \citep{Fiedler:1987ApJS,Waltman:1991ApJS,Lazio:2001ApJS}. Starting from the PDFs we build the cumulative distribution functions (CDFs) of $\theta_N$ and $m_N$ (Fig. \ref{ISM_mr_theta}). For $\nu_s=8$ GHz the agreement is generally good; however, the model underestimates the lower end of the distributions. Such discrepancy is likely to arise from the observational uncertainties in the flux density calibration. This translates into a lower limit for the modulation, $m_N>0.01$. 
We also estimate an uncertainty of $\sim 100\, \mu$as on $\theta_{N}$ since angular measurements are inferred using a flux dependent model~\citep{Rickett:2006}.

For the same numerical set-up, we have also allowed the source frequency to vary in the range $1-50$ GHz, both for validation purposes and also as a diagnostic to isolate the ISS contribution from the IGM one. As we can see in Fig. \ref{ISM_source_overview_fig}, both $\theta_N$ and $m_N$ decrease for $\nu_{s}\ge\nu_{c}\simeq2\mbox{ GHz}$. The critical frequency $\nu_c$ marks the transitions from the diffractive ($\nu<\nu_c$) to the refractive ($\nu>\nu_c$) regime. The exact value of $\nu_{c}$ depends on the properties of the screens through the refractive condition given in eq. \ref{condizionerifrattiva_screens}. This is consistent with \citet{Rickett:1990}, in which the diffractive regime start below $\sim$ 5 GHz. Hence, in the following we restrict our predictions to $\nu_{s}>5$ GHz.

For a medium characterized by Kolmogorov turbulence in the refractive regime eqs. \ref{eq_thetaf} and \ref{m_r-goodman} apply, and therefore, for weak scintillation ($\theta_{F}\gg\theta_{d}$), $\theta_{N}\propto\nu_{s}^{-1/2}$, and $m_{N}\propto \nu_{s}^{-31/12}$. The numerical results of Fig. \ref{ISM_source_overview_fig} are in close agreement with these analytical predictions. Note that a longer $T_{\obs}$ does not affect the results, since (see eq. \ref{angolorifrazionetotale_1}) $t_{\refr}< 1$ d over the entire frequency range considered.

Having shown that our model can reliably explain the observed properties of ISS, we can extend our analysis to the scintillation of extra-galactic sources.

\section{Cosmological simulations}\label{ref_sezione_cosmo}

\begin{figure}
\begin{center}
\ifpdf
  \includegraphics[width=8.3cm]{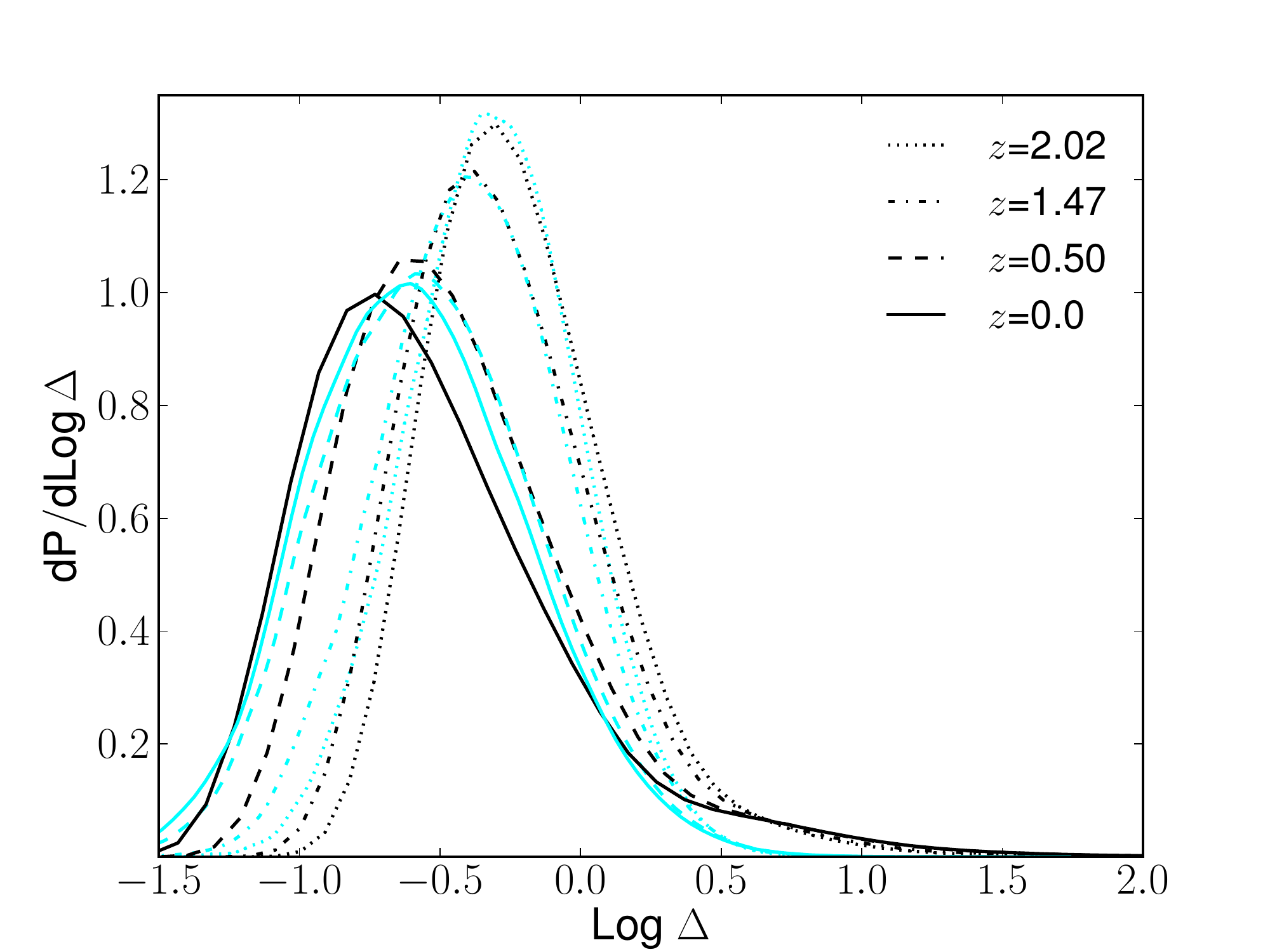}
\else
  \includegraphics[width=8.3cm]{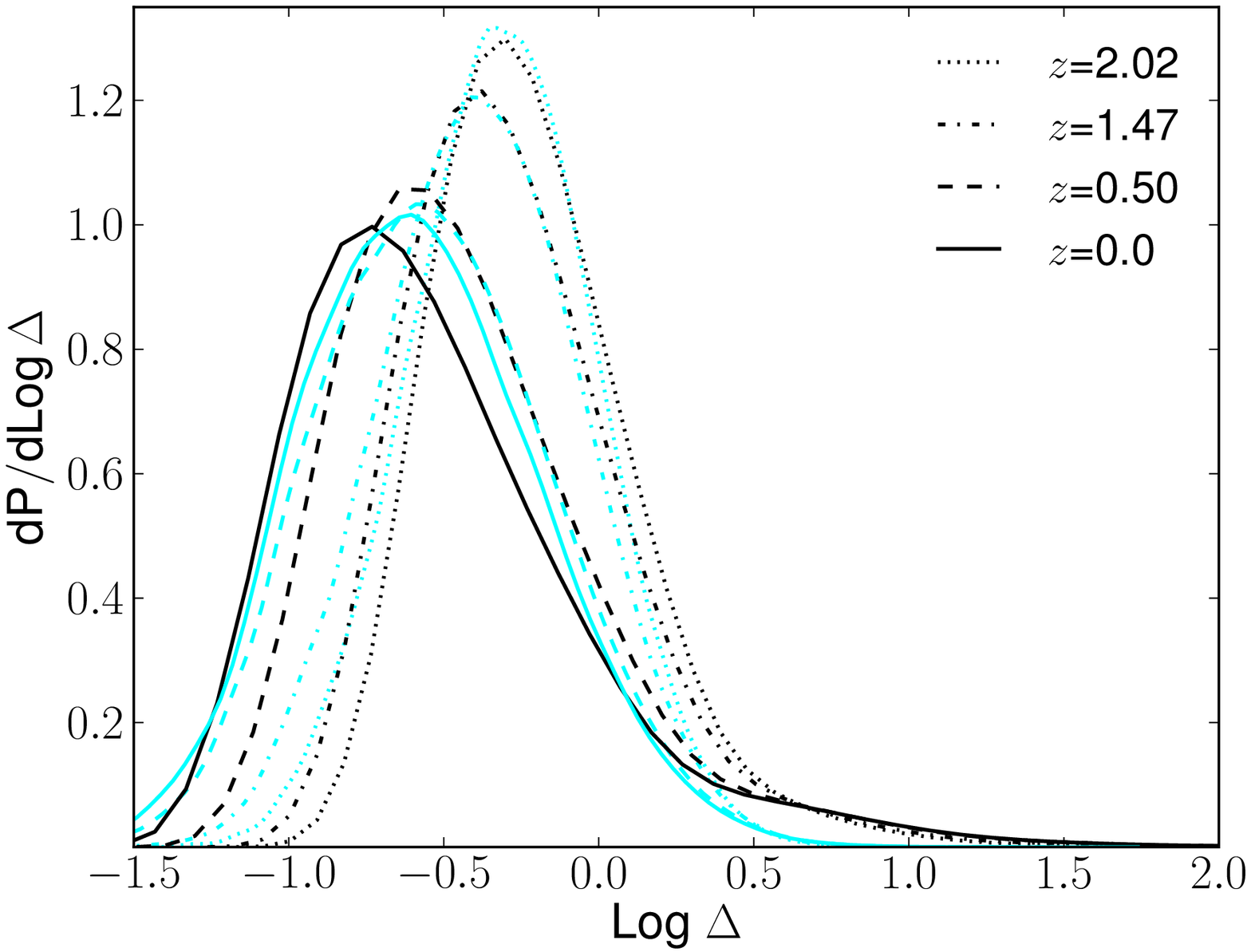}
\fi
\end{center}
\caption{Volume-weighted PDF of the baryon overdensity for redshifts $0 \leq z \lsim 2$, obtained from simulations (black lines) and log-normal distribution (cyan).\label{n_vs_rho_fig}}
\end{figure}
For the present study we have performed cosmological simulations using the publicly available code {\tt RAMSES}~\citep{Teyssier:2002}, which can be described as a Fully Threaded Tree (FTT) data structure where the hydrodynamical Adaptive Mesh Refinement (AMR) scheme is coupled with a Particle Mesh (PM3) N-body solver employing a Cloud-in-Cell interpolation scheme to solve the Poisson equation.
\begin{figure*}
\begin{center}
\ifpdf
  \includegraphics[width=8.3cm]{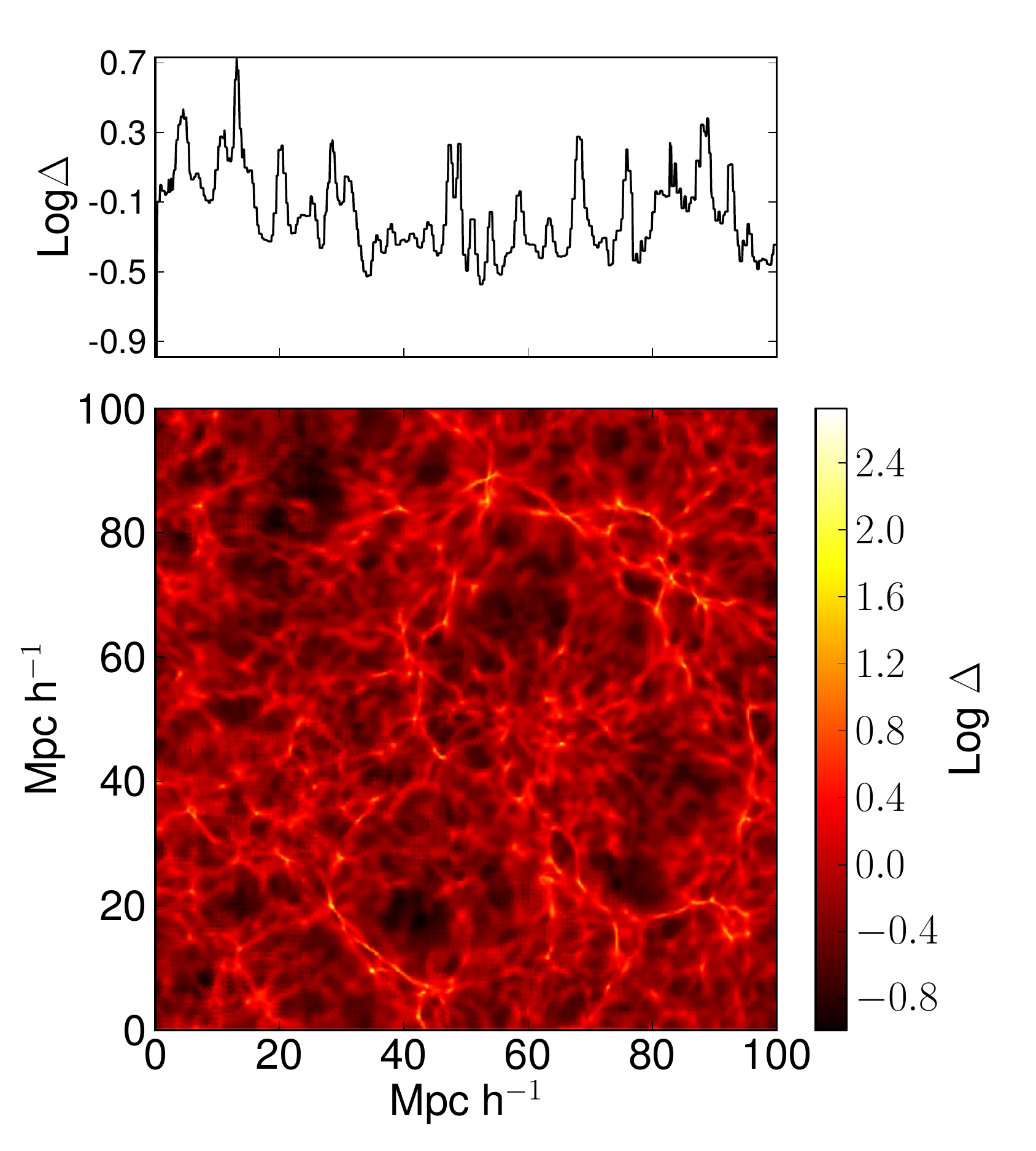}
  \includegraphics[width=8.3cm]{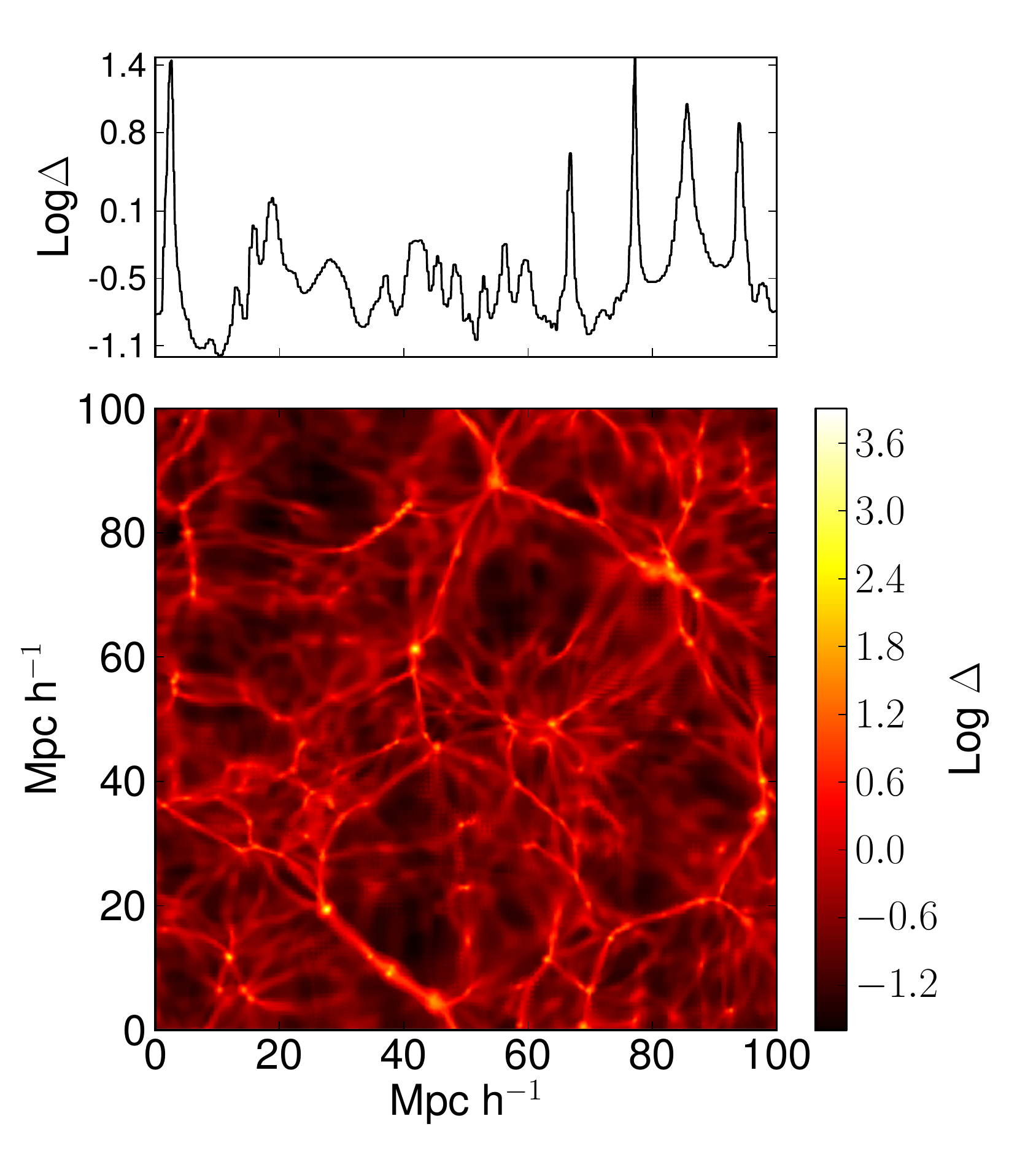}
\else
  \includegraphics[width=8.3cm]{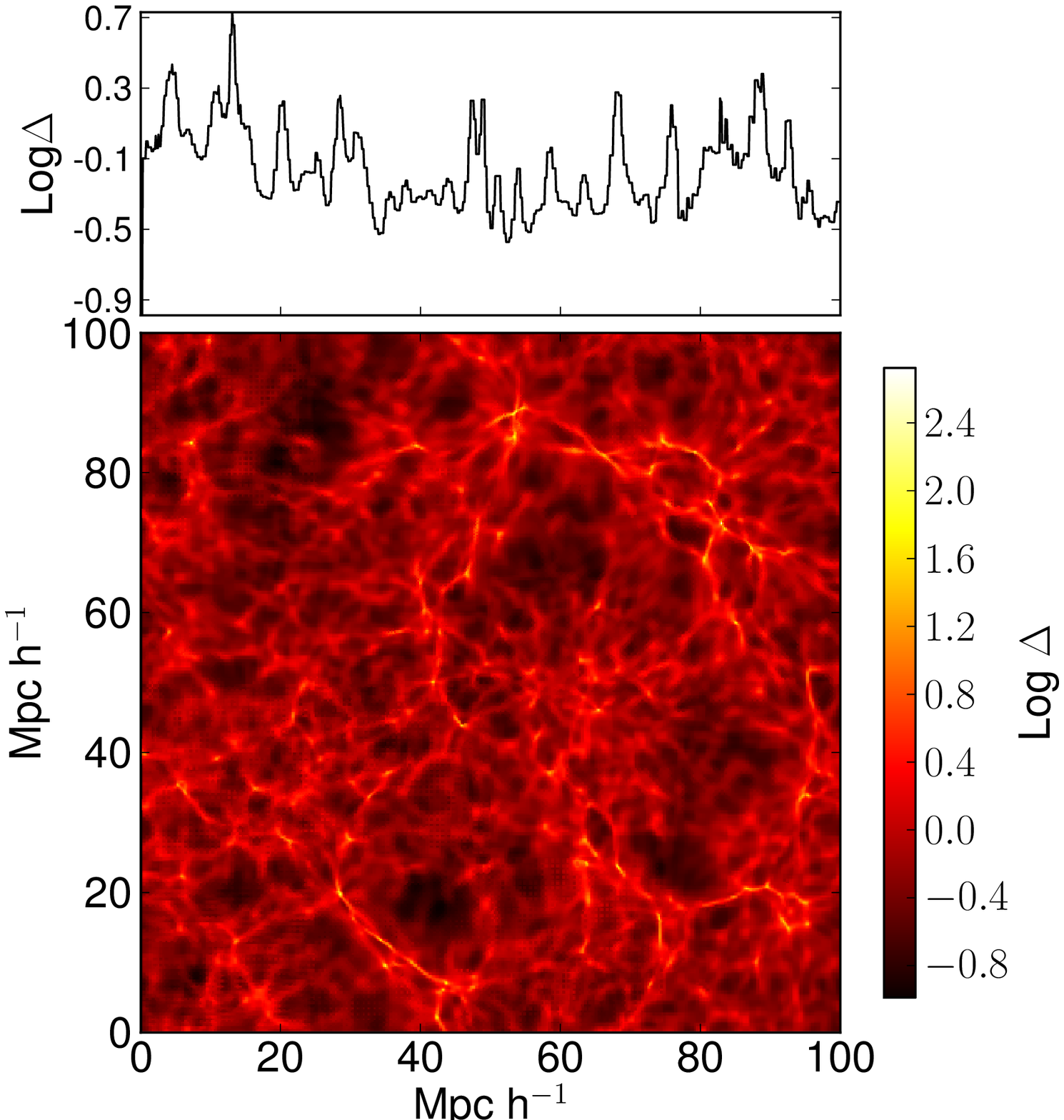}
  \includegraphics[width=8.3cm]{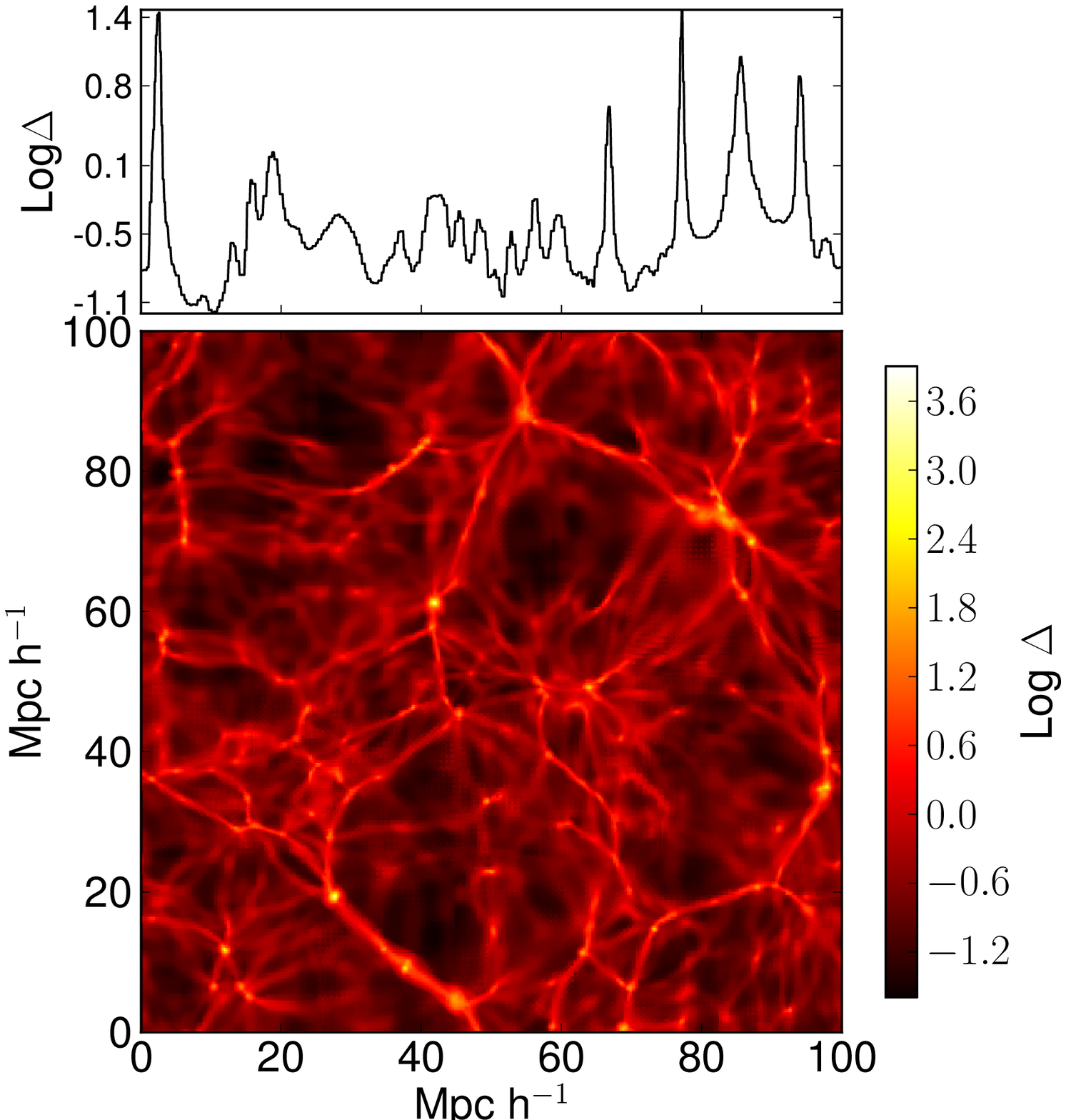}
\fi
\end{center}
\caption{\label{densita_fig_3_0}
{\it Lower panels}: IGM density field, $\Delta=\rho\slash\langle\rho\rangle$, at $z=3$ (left) and $z=0$ (right) in a slice through the simulation box of thickness $0.3\ h^{-1}$ Mpc.
{\it Upper panels}: 1D density cut along the horizontal line at 50 Mpc $h^{-1}$ through the map below.
}
\end{figure*}

The simulation box size of $100\,h^{-1}$ comoving Mpc is resolved with $256^{3}$ dark matter particles; we allow 6 additional levels of refinement for the baryonic matter using the canonical cosmological refinement strategy in which a cell is resolved with a finer grid if its density is $>8$ times the mean. This yields a mass resolution of $1.65 \times 10^{10} \ \Omega_{dm}h^{-1}\ M_{\odot}$ for the dark matter and a formal spatial resolution of $6\,h^{-1}$ kpc for baryons.
The simulation starts at $z=100$ with initial conditions generated using the {\tt GRAFIC} routine~\citep{Bertschinger:2001}. Given the intrinsic statistical uncertainties in the approach we use to model the IGM scintillation, our results are essentially insensitive to changes of the cosmological parameters within 1-$\sigma$ c.l..
\begin{figure*}
\begin{center}
\ifpdf
  \includegraphics[width=8.3cm]{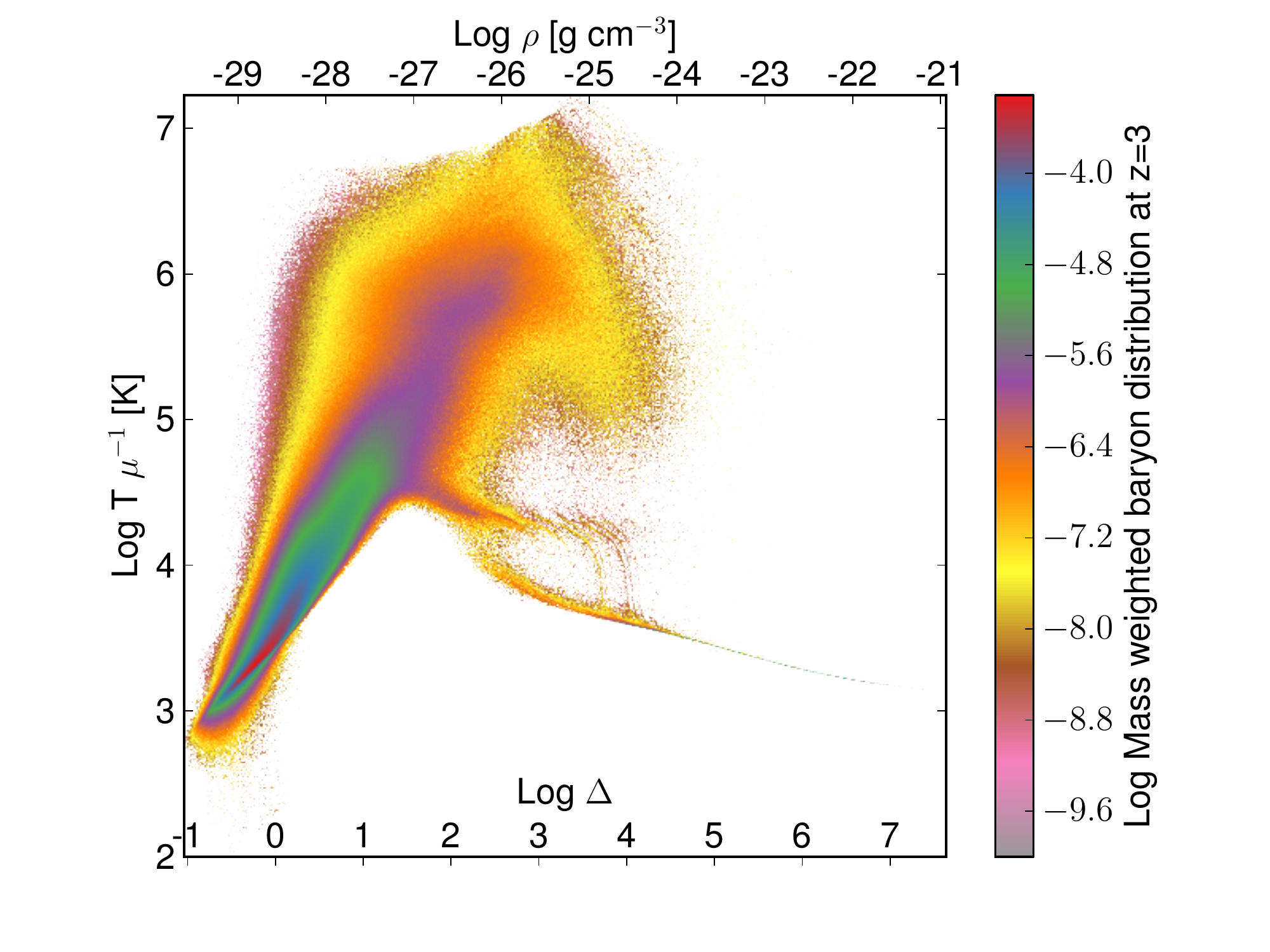}
  \includegraphics[width=8.3cm]{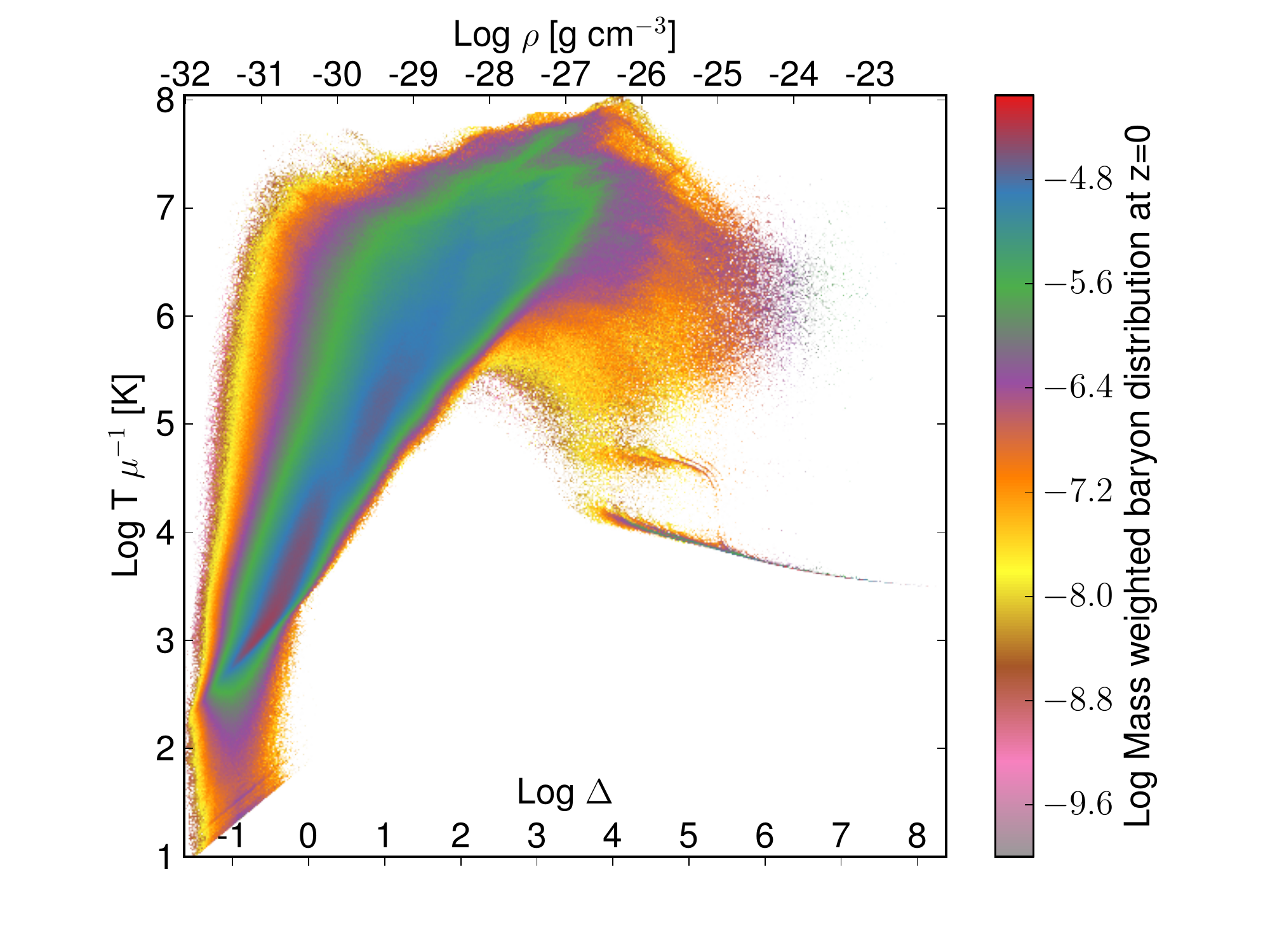}
\else
  \includegraphics[width=8.3cm]{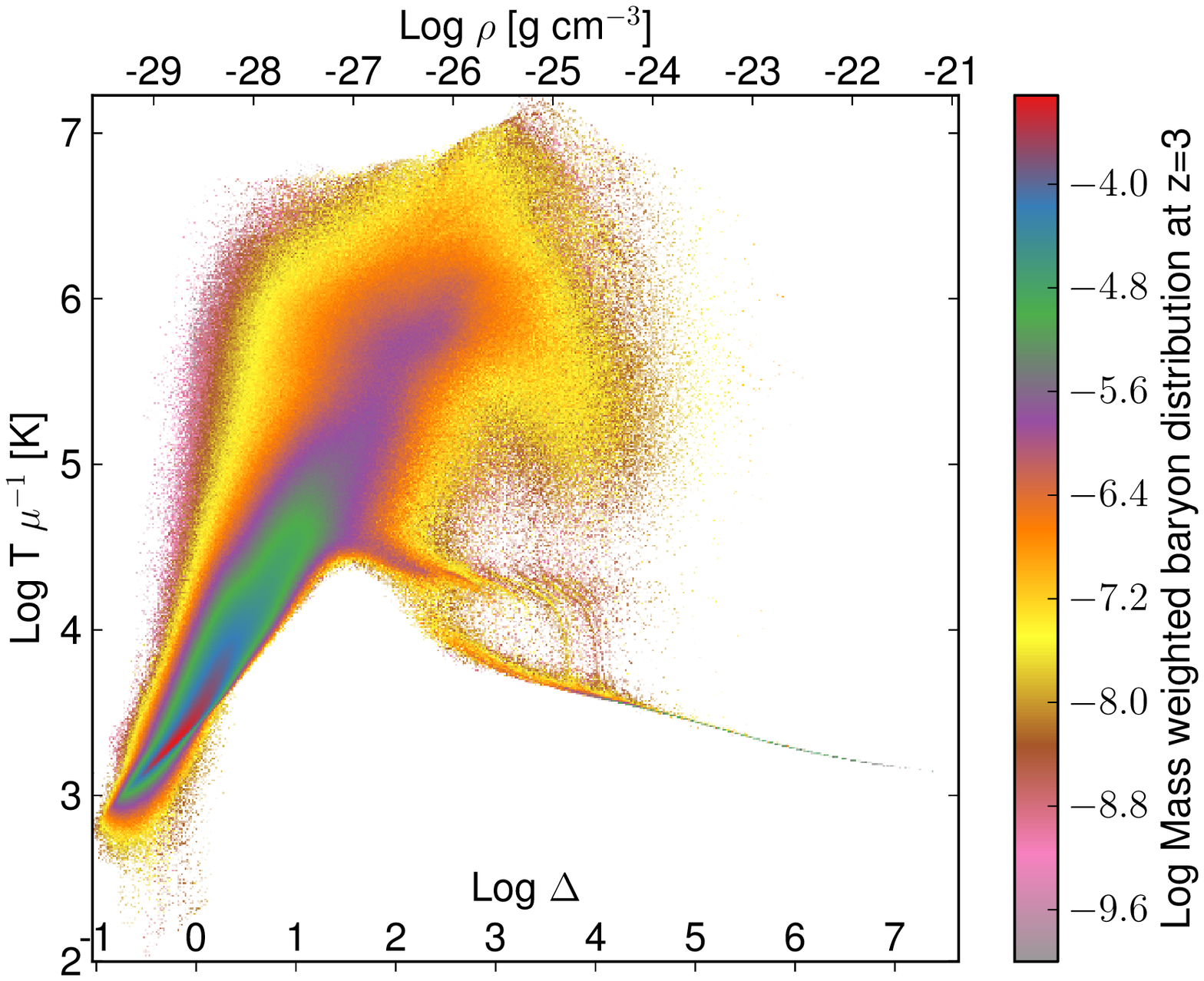}
  \includegraphics[width=8.3cm]{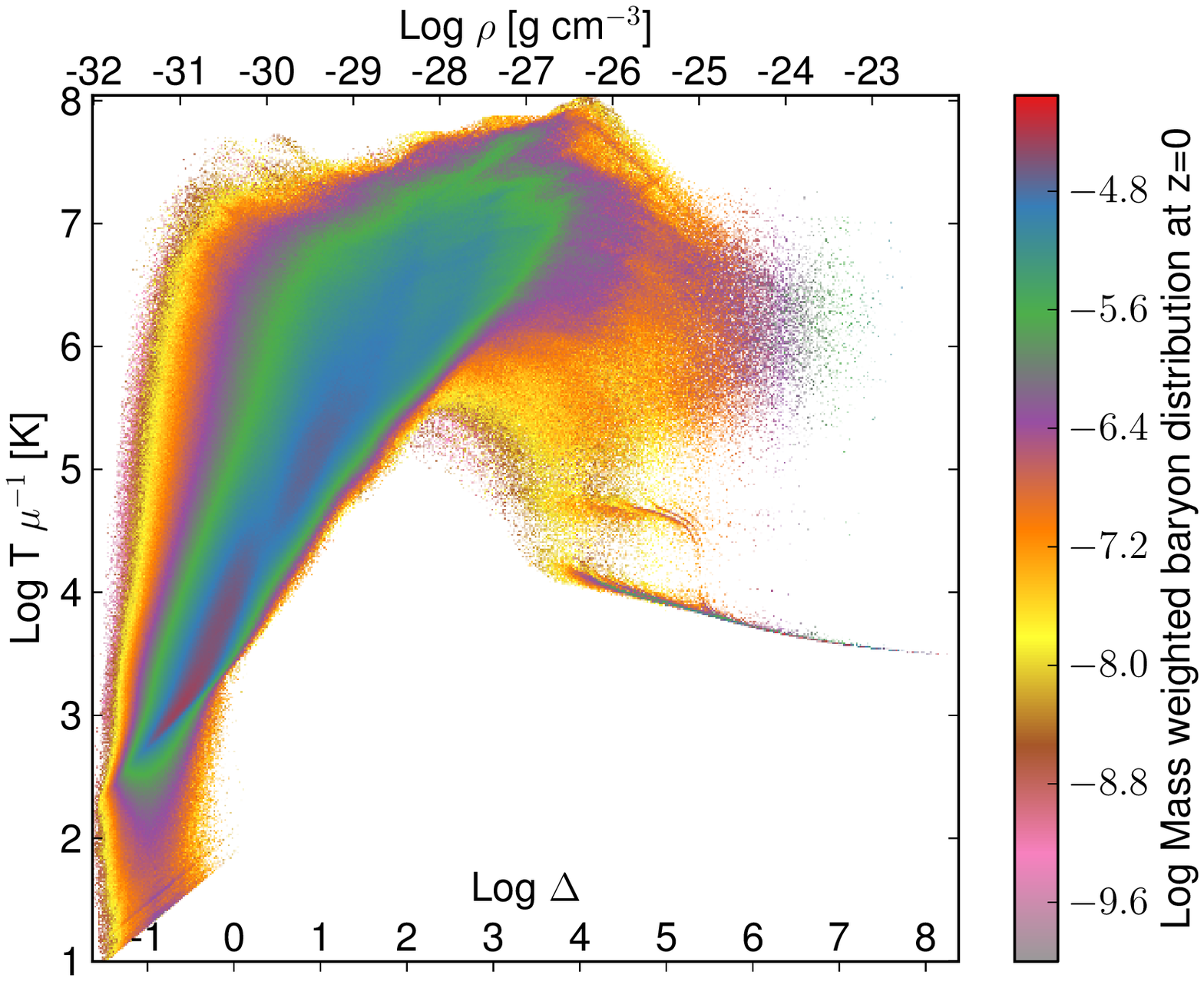}
\fi
\end{center}
\caption{IGM equation of state at $z=3$ (left) and $z=0$ (right). The colorbar represents the differential mass-weighted probability function; temperature is expressed in molecular weight units.\label{stato_fig_03}}
\end{figure*}

Star formation is not included in the simulation as sub-galactic scales are poorly resolved; in addition, a given l.o.s. has negligibly low probability to intersect a star forming region. Similarly we have not attempted to model large scale effects of IGM turbulence since (i) the assumed turbulent scale responsible for the scintillation is $< 1$ pc \citep[][]{Evoli:2011} and (ii) they do not strongly affect the thermal/ionization state of the gas \citep{Iapichino:2013}.

The heating-cooling processes in {\tt RAMSES} are handled using the moment-based radiative transfer code {\tt ATON} \citep[][]{Aubert:2008}, which includes the UV background of \citet{Haardt:1996, Haardt:2012} given by stellar+quasar contributions \citep{Haehnelt:2001}, which has been explicitly tested in \citet{Theuns:1998MNRAS}. Taking into account this contribution is important in determining the ionized fraction which in turn is fundamental to correctly calculate the scattering measure which depends of the free electron density.

The inferred gas density distributions, expressed in terms of $\Delta \equiv \rho\slash\langle\rho\rangle$, are shown for different redshifts by black lines in Fig.~\ref{n_vs_rho_fig}. The PDFs are consistent with the results by~\citet{Miralda-Escude:2000}, which in turn are calibrated against Gunn-Peterson constraints. Additionally, we have compared the simulated PDFs with the ones obtained from a log-normal model (LNM). The LNM \citep[i.e.][]{Coles:1991MNRAS,Choudhury:2001MNRAS} is a semi-analytical model embedding the first order corrections to the linear evolution of the baryon overdensity field; this is obtained by filtering the dark matter density with the baryonic Jeans length \citep{Bi:1997ApJ,Gnedin:2000ApJ}. The comparison is shown in Fig. \ref{n_vs_rho_fig}. The LNM reproduces the simulated baryonic PDF but fails to match the high $\Delta$ tail. In addition the overall agreement degrades towards low redshift ($z\lsim0.5$). We will use the LNM for a comparison with the simulated IGM scintillation predictions, to distinguish the contribution from largely non-linear overdensities.

The simulated gas density field, featuring the typical cosmic web structure made of filaments and density knots corresponding to
galaxy groups and proto-clusters, is shown in Fig.~\ref{densita_fig_3_0} for $z=3$ and $z=0$. In addition to the spatial information, we can also analyze the IGM thermodynamic properties in term of the equation of state. Three distinct phases can be identified: (i) a photo-ionized phase, i.e. the so-called \Lya forest, characterized by relatively low densities, $\Delta\lesssim 10^{2}$, and temperature $T\mu^{-1}\lesssim 10^{5}$ K; (ii) a shock-heated phase with $T >10^{5}$ K (WHIM); and (iii) a cold phase ($T\mu^{-1}\lesssim 10^{4}$ K), made of dense ($\Delta\gtrsim 10^{4}$) pressure-supported clumps that can host star formation, slowly built by structure formation (Fig. \ref{stato_fig_03}).

These results are broadly consistent with those form similar studies in the literature. For example, \citet{Peeples:2010} simulated a $12.5 h^{-1}$ Mpc comoving box with an high ($2\times288^{3}$ particles) and low ($2\times144^{3}$ particles) resolution using the Smooth Particle Hydrodynamic (SPH) code {\tt GADGET-2} \citep{Springel:2005MNRAS}. At $z=3$ they found an equation of state (Fig 1. of their paper) very similar to the one obtained here (Fig. \ref{stato_fig_03}); however, there are differences at the highest densities, which are related to the different box size, resolution and intrinsic differences between AMR and SPH. At $z=0$ we compare the results with \citet{Rasera:2006} who performed a convergence test on an extended set of {\tt RAMSES} simulations; their equation of state (Fig. 2 of their paper) is consistent with the one shown in Fig.~\ref{stato_fig_03}.
\section{Results}\label{ref_sezione_analisi}
The analysis of the IGM scintillation of extra-galactic sources is performed considering $N_{\los}$ l.o.s.. We start by evaluating the simulated SM up to $z=2$. Then we apply the full numerical scheme described to compute the expected scintillation properties of distant sources at given $\nu_s$ and $z$. Finally, we perform an exploration of the $(\nu_s,z)$ parameter space.
\subsection{Intergalactic scattering measure}

Using eq. \ref{eq_def_smequ}, a simple estimate of the equivalent $\mbox{SM}$ for a smooth Friedmann-Robertson-Walker universe can be written as:
\begin{equation*}
	\mbox{SM}_{\frw}=\overline{\mbox{SM}}_{-3.5}\int_{a_{s}}^{1}\frac{\mbox{d}a}{a^{7}E(a)}\, ,
\end{equation*}
with
\begin{equation*}
\left\{
\begin{aligned}
E(a)&=\sqrt{\Omega_{\lambda}+\Omega_{r}a^{-4}+\Omega_{m}a^{-3}}\\
\overline{\mbox{SM}}_{-3.5}&=\left[\frac{\rho_{c}\Omega_{b}\left(m_{p}\mu\right)^{-1}}{0.02\mbox{ cm}^{-3}}\right]^{2}\frac{c\,\slash H_0}{\kpc}\, ,
\end{aligned}
\right.
\end{equation*}
where $m_{p}$ is the proton mass, $\mu$ the mean molecular weight and we have approximated the group velocity of the wave in the medium to $\approx c$, which is reasonable for the frequency range we are considering. Then the mean SM for a source at $z=2$ in the assumed $\Lambda$CDM model, for a fully ionized IGM is
\begin{equation}\label{meanSM}
\mbox{SM}_{\frw} \simeq 0.0984\, .
\end{equation}
\begin{figure}
\begin{center}
\ifpdf
  \includegraphics[width=8.3cm]{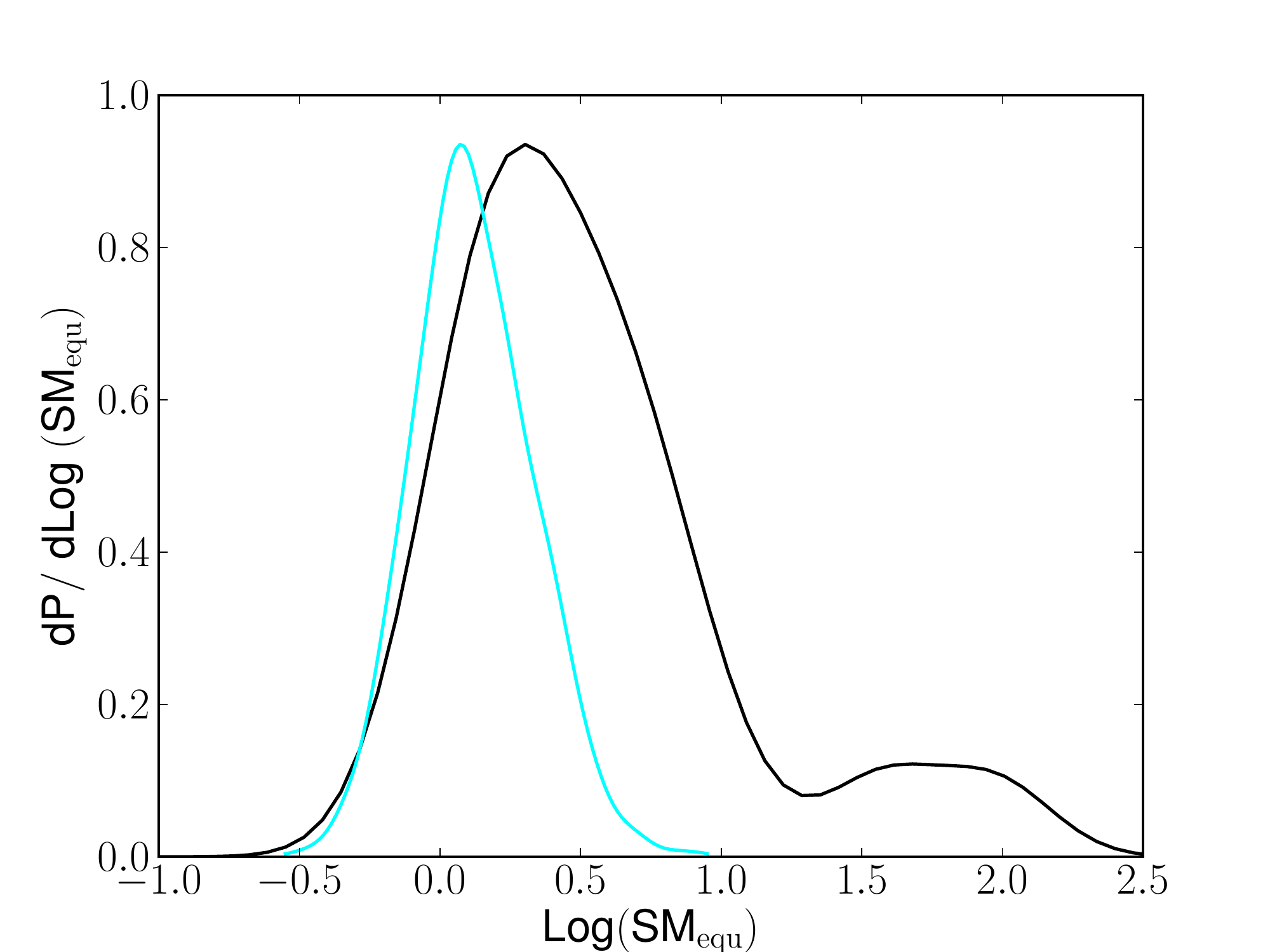}
\else
  \includegraphics[width=8.3cm]{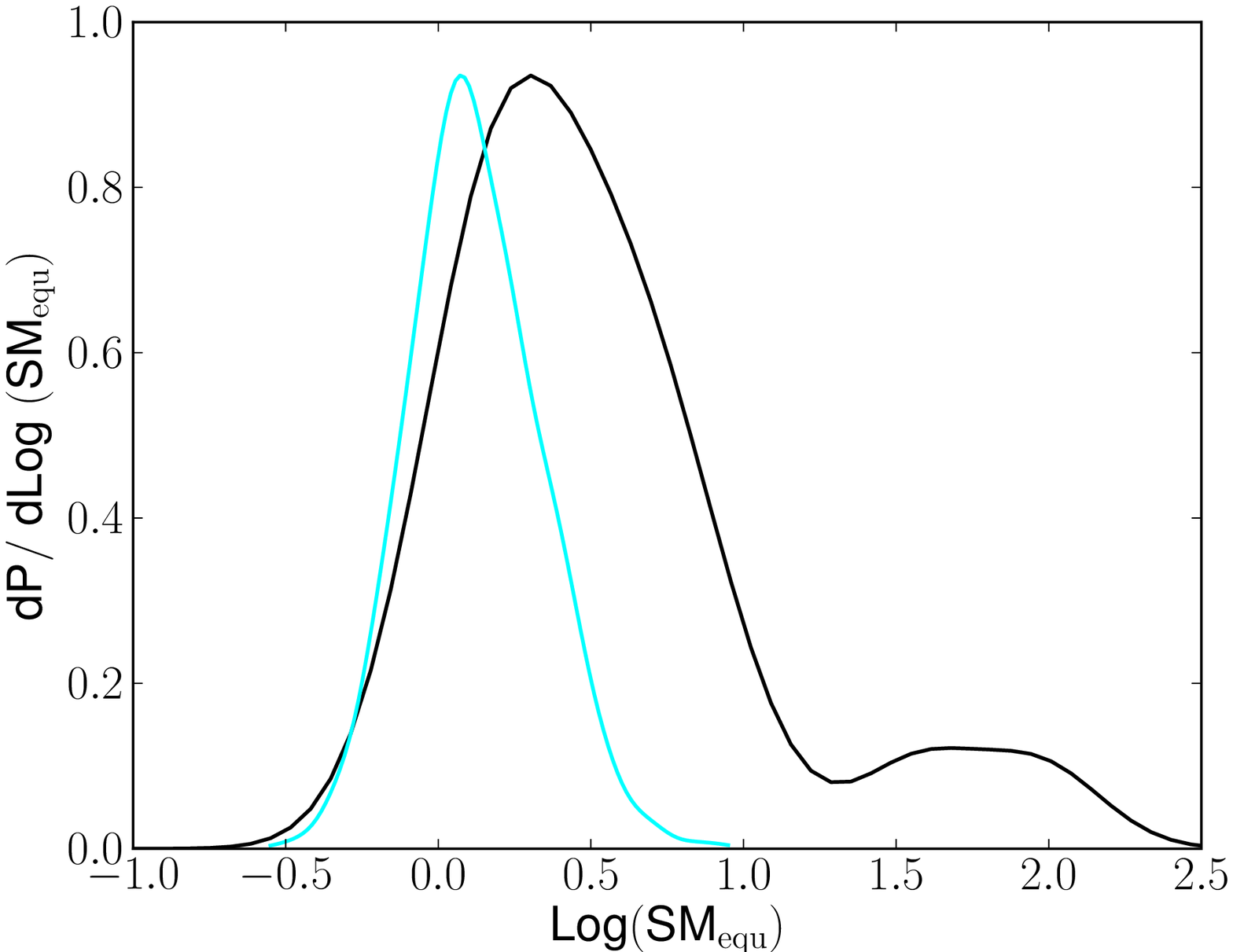}
\fi
\end{center}
\caption{IGM equivalent scattering measure PDF for a source located at $z=2$. The PDF obtained from the cosmological simulation (LNM) is shown by the black (cyan) line. The simulated PDF has a mean $\langle \mbox{SM}_{\equ}\rangle=3.879$ and a r.m.s. $\sigma=8.612$. The LNM yields $\langle \mbox{SM}_{\equ}\rangle=1.318$ and $\sigma=1.151$.
\label{IGM_SM_fig}}
\end{figure}
In Fig.~\ref{IGM_SM_fig} with a black line we show the SM$_{\equ}$ PDF for sources at $z=2$. The mean equivalent SM is $\langle \mbox{SM}_{\equ}\rangle=3.879$, i.e. almost 40 times larger than for a smooth IGM (eq. \ref{meanSM}). As $\mbox{SM}\propto \rho^{2}$, the scattering measure is very sensitive to density fluctuations induced by the gravitational instability. Thus the value of eq. \ref{meanSM} is grossly inaccurate and represents only a lower limit to the actual SM.

A better estimate can be obtained\footnote{For display purposes the PDFs inferred from the LNM are plotted without errors and normalized to have the same maximum values as the simulated ones.} by using the LNM (Fig. \ref{IGM_SM_fig}). From there we see that the mean $\langle\mbox{SM}_{\equ}\rangle$ of the two distributions are comparable, but the LNM gives a much smaller variance and does not show the tail of large SM$_{\equ}$ values. 

\subsection{Single monochromatic source}
\begin{figure}
\begin{center}
\ifpdf
  \includegraphics[width=8.3cm]{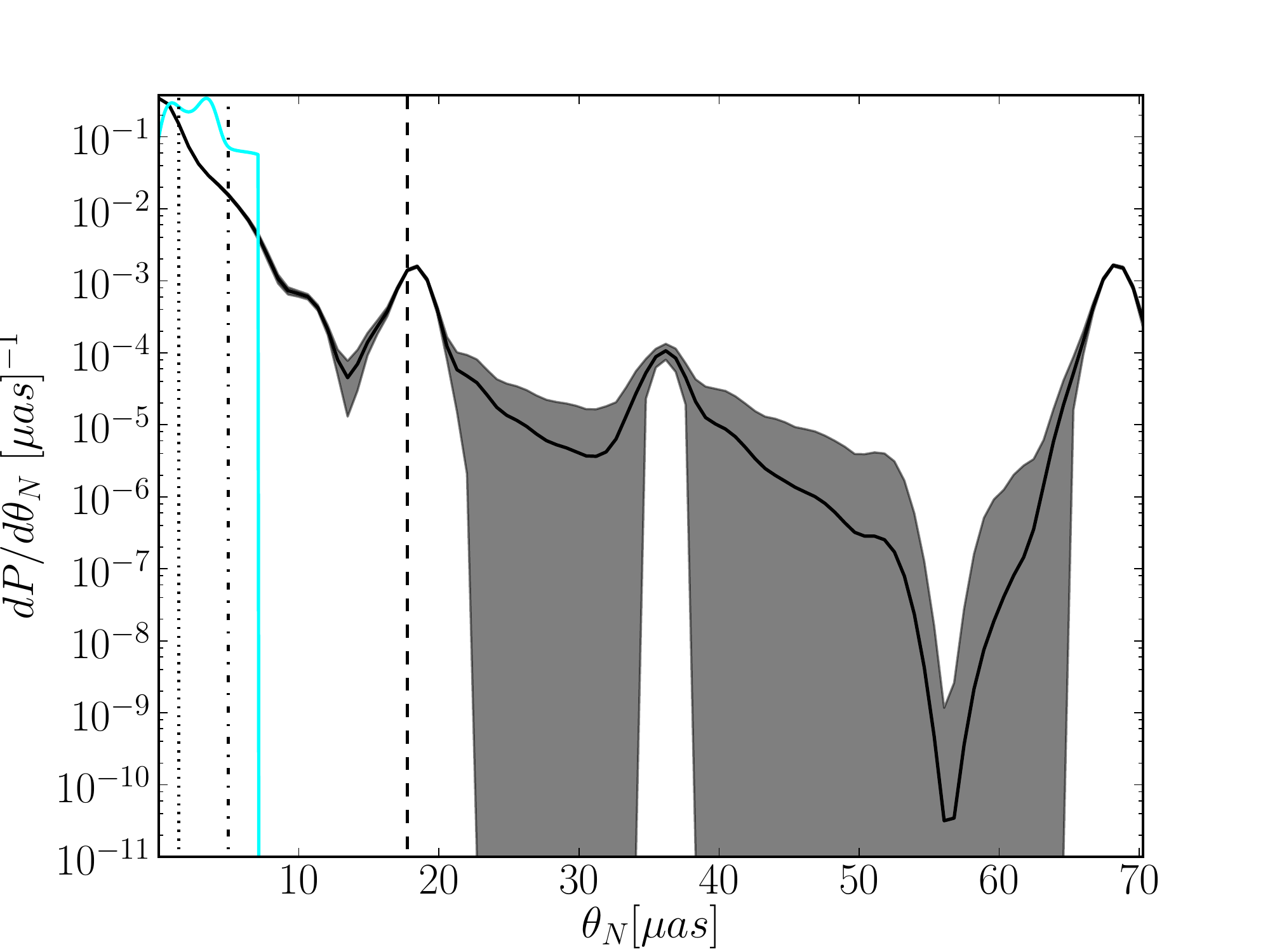}
\else
  \includegraphics[width=8.3cm]{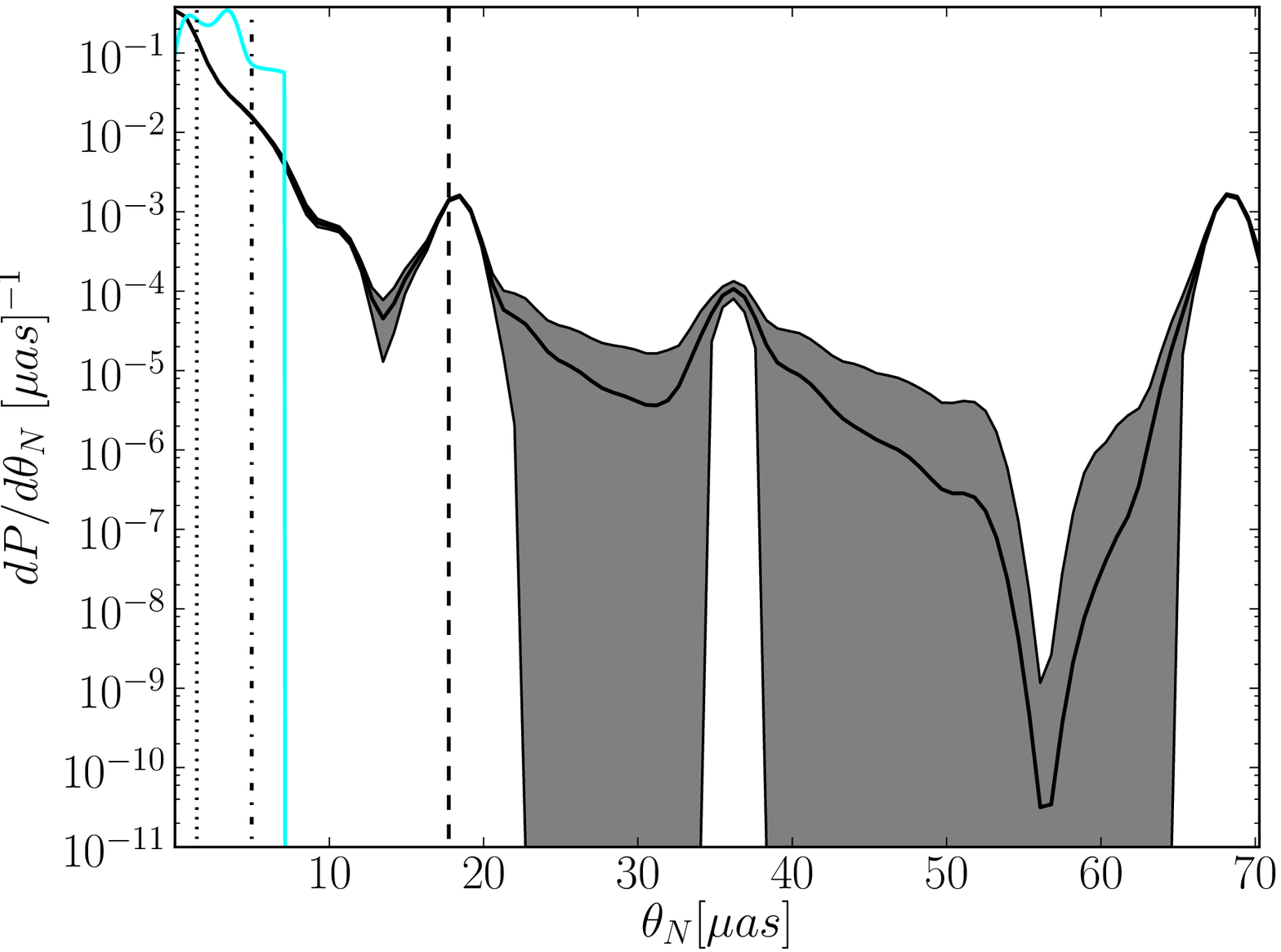}
\fi
\end{center}
\caption{IGM refraction angle PDF for a source located at $z=2$ at restframe frequency $\nu_{s}=5$ GHz. The solid black line is the PDF from the cosmological simulation, the shaded region indicates its error (see Appendix \ref{app_PDF}) and the cyan solid line is the PDF inferred from the LNM. The simulation PDF has a mean value $\langle\theta_{N}\rangle$=$(1.77 \pm 0.11)\,\mu$as and the r.m.s. is $\sigma$=$(5.58\pm1.95)\,\mu$as. The LNM PDF has $\langle\theta_{N}\rangle$=$(2.74\pm0.23)\, \mu$as $\sigma$=$(1.66\pm 0.45)\,\mu$as. For the simulation only, vertical lines identify values of $\theta_N$ whose cumulative probability is $(0.68,\ 0.95,\ 0.99)$, from left to right respectively.\label{IGM_theta_8_fig}}
\end{figure}

To better understand the properties of IGM scintillation, we focus our attention to a typical case in which a source located at $z=2$, is emitting at the restframe frequency $\nu_{s}=5$~GHz for an observing $T_{\obs}=2$~d. This choice of parameters will enable in the next Section a direct comparison of our results and recent observational data for extra-galactic sources.

\begin{figure}
\begin{center}
\ifpdf
  \includegraphics[width=8.3cm]{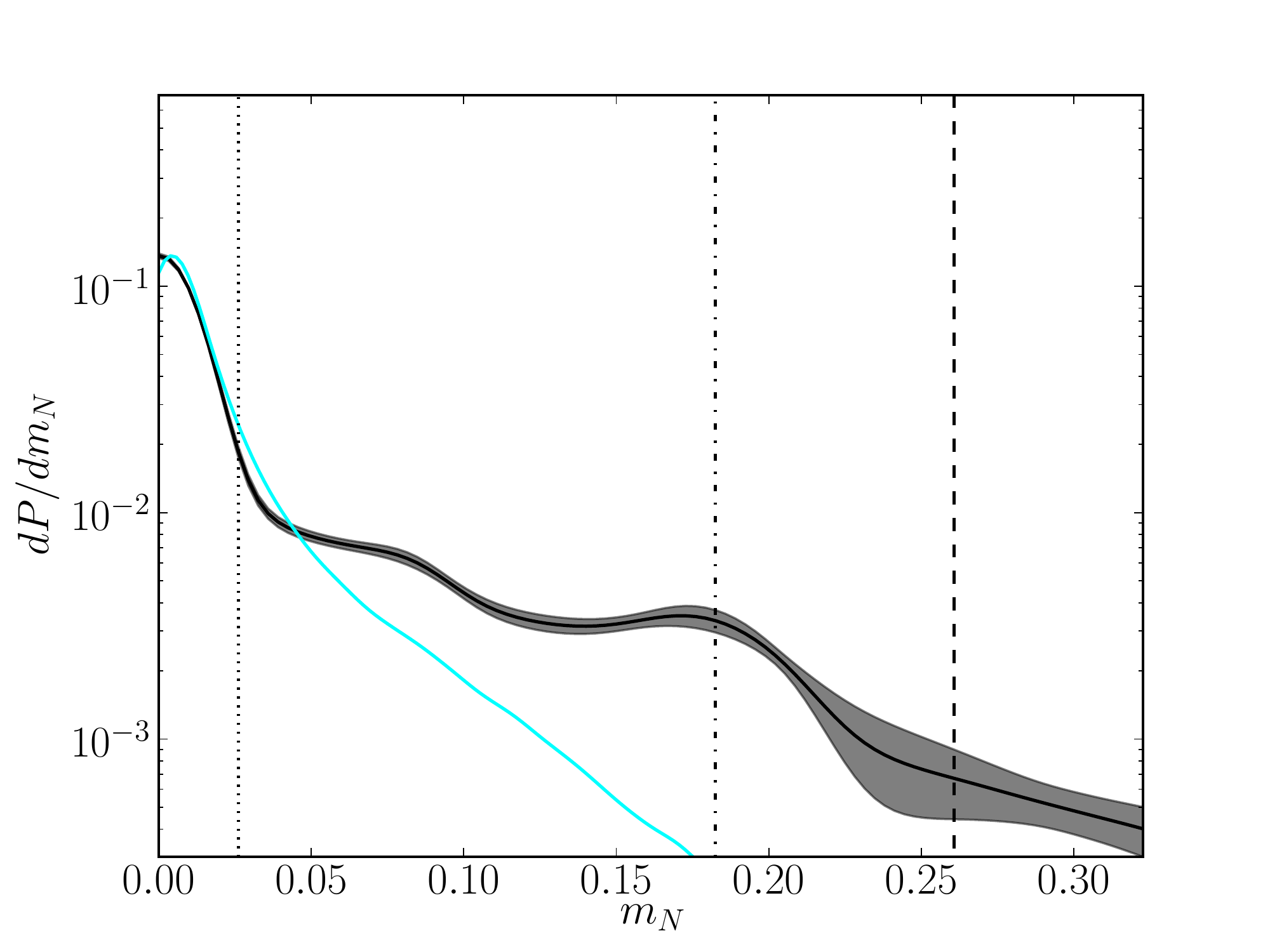}
\else
  \includegraphics[width=8.3cm]{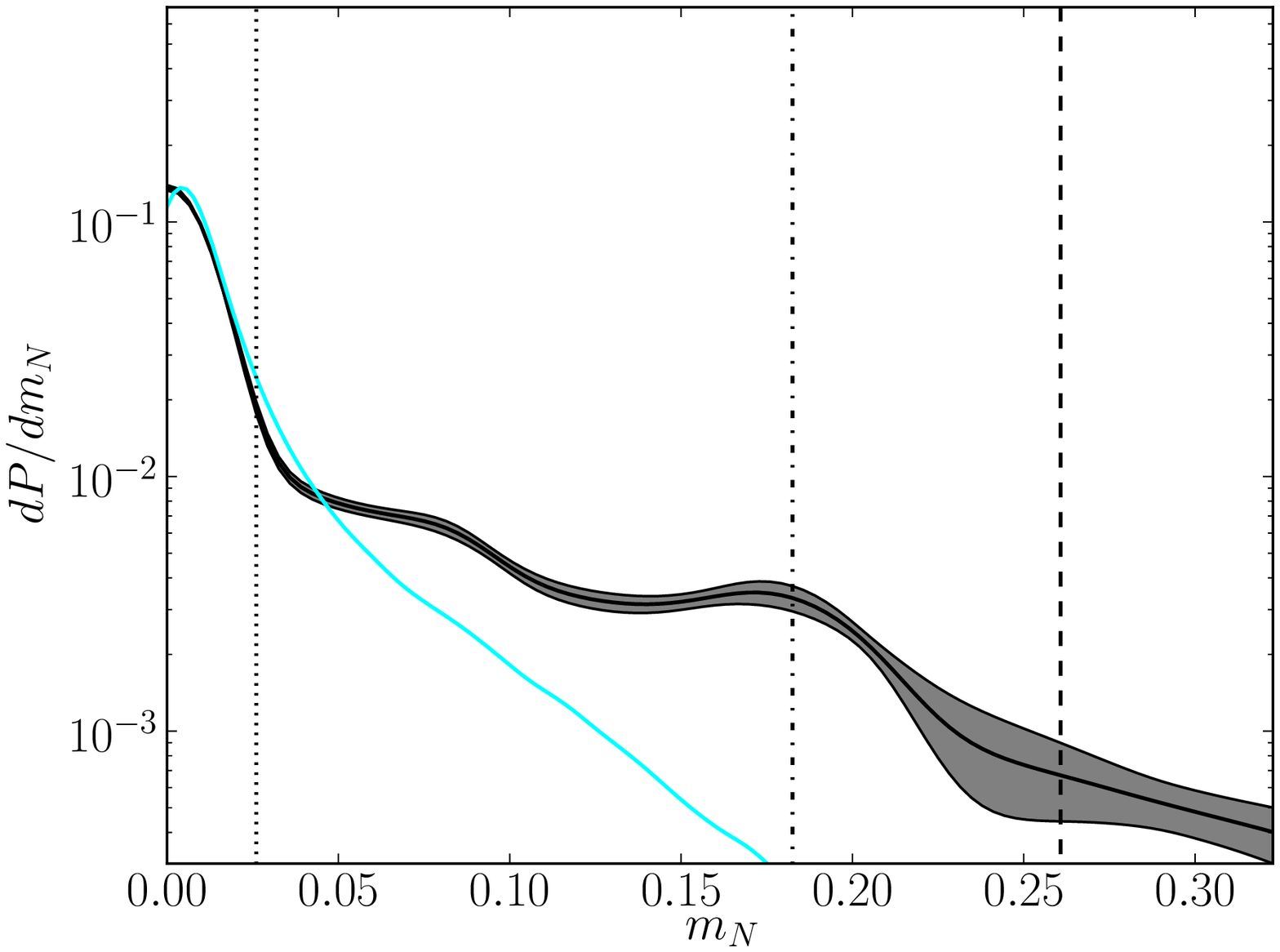}
\fi
\end{center}
\caption{As Fig. \ref{IGM_theta_8_fig} for the modulation index. The simulation PDF has mean value $\langle m_{N}\rangle$=$0.0389 \pm 0.0037$  and r.m.s. $\sigma$=$0.060\pm0.022$. The LNM PDF has $\langle m_{N}\rangle$=$0.018\pm0.002$ and $\sigma$=$0.025\pm0.011$.\label{IGM_m_r_8_fig}}
\end{figure}

In Fig.~\ref{IGM_theta_8_fig} and Fig.~\ref{IGM_m_r_8_fig} we present our results for the the refraction angle and the modulation index respectively. Compared to the cosmological simulation, the LNM yields PDFs with comparable mean values but much steeper decreases (or even a sharp cut in the case of the refraction angle) towards large values of $\theta_N$ and $m_N$. This is expected on the basis of the previous SM$_{\equ}$ comparison, and strongly suggests an interpretation in which large refraction angle and/or modulation indexes can only be produced in l.o.s. passing near or through highly non-linear or virialized structure.

The PDF of the refraction angle (Fig. \ref{IGM_theta_8_fig}) has a general power-law shape to which several peaks are superposed; these corresponds either to l.o.s. passing through a single large overdensity (i.e. a proto-cluster) or to the coherent sum of smaller ones. This is in line with the expectation from a Levy flight distribution. The probability to obtain a $\theta_N$ within 1-$\sigma$ of the mean is $P\left(\left| \theta_{N}-\langle \theta_{N}\rangle \right| < \sigma\left(\theta_{N}\right) \right) = 0.979 \pm 0.008$; since $\sigma=(5.58 \pm 1.95)\,\mu$as some l.o.s. could yield a refraction angle much larger than $\langle\theta_{N}\rangle$. While the IGM 1-$\sigma$ angle is much smaller than the corresponding one due to ISM scintillation (see Fig. \ref{ISM_source_overview_fig}), its value is more than 7 times the one inferred for a smooth IGM \citep[i.e.][]{2007mru..confE..46R}. This has interesting consequences, as we will discuss below.

The modulation index distribution (Fig. \ref{IGM_m_r_8_fig}) shows a significantly different trend. After a steep decline, $\mbox{d}P/\mbox{d}m_N$ flattens and stabilizes to $\approx 10^{-3}$ before a final decrease for up to $m_N \simgt 0.2$. The peaks caused by large intersected overdensities are still visible, although less pronounced than in the case of the refraction angle. The 1-$\sigma$ probability is $P\left(\left| m_{N}-m\left(m_{N}\right) \right| < \sigma\left(m_{N}\right) \right) = 0.864 \pm 0.022$, and the $m_{N}$ value is comparable to ISS at the same frequency (Fig. \ref{ISM_source_overview_fig}).

This fact can be understood as a result of two competing effects (see eq. \ref{m_r-goodman}): (i) $m_N$ depends on the scattering as $m_r\propto \mbox{SM}^{1\slash 2}$, but (ii) it has only a weak dependence on the screens distance ($m_r\propto \tilde{d}^{-1\slash 6}$). While from (i) we would expect a sub-dominant contribution of ISS to $m_{r}$, the distance dependence implied by (ii) enhances the role of ISS. However, for the frequency range of interested here, the net result 
is that the IGM contribution to $m_r$ can compete and possibly overcome the ISM one.
\subsection{Redshift and frequency dependence}
\begin{figure}
\begin{center}
\ifpdf
  \includegraphics[width=8.3cm]{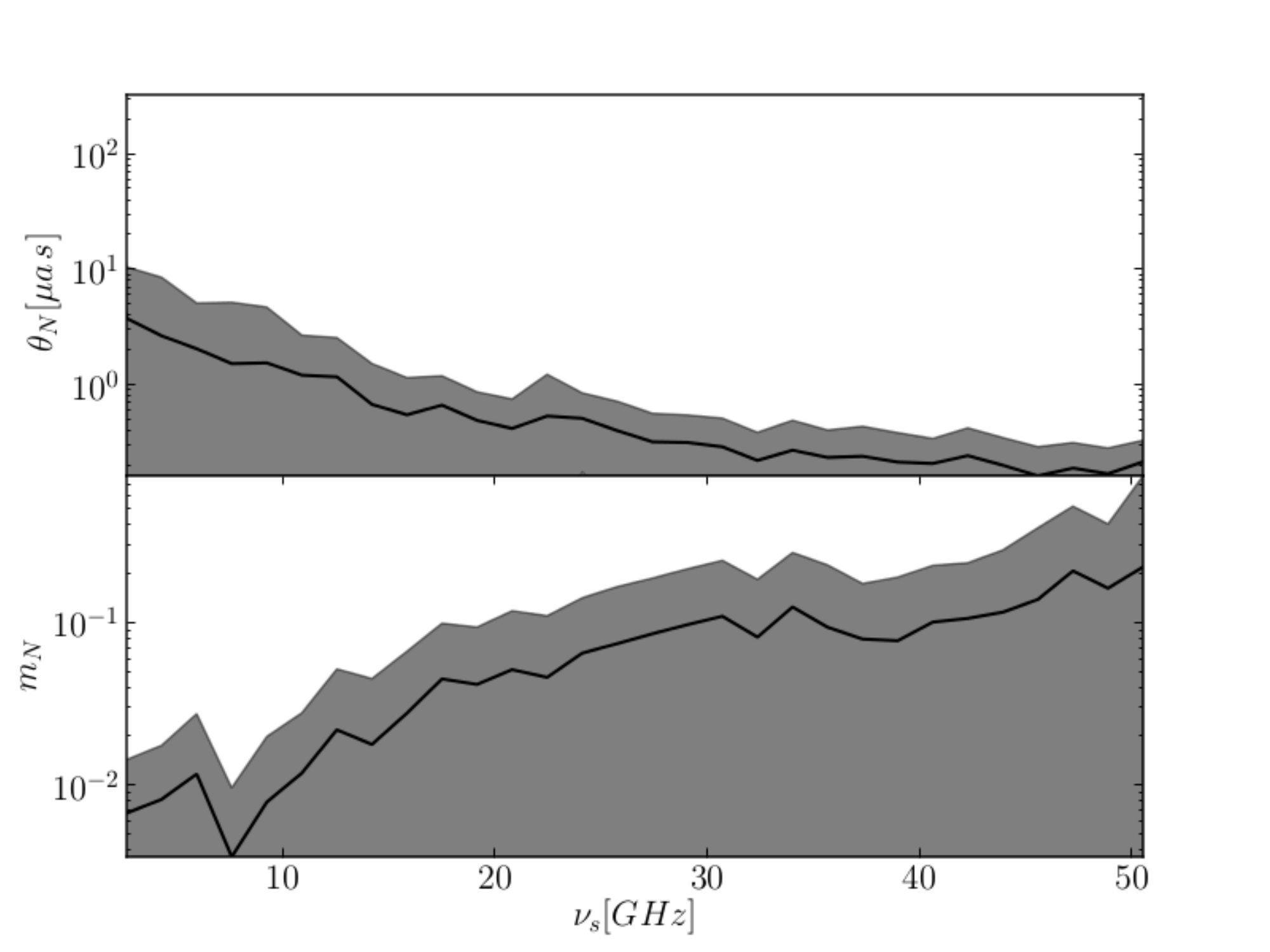}
\else
  \includegraphics[width=8.3cm]{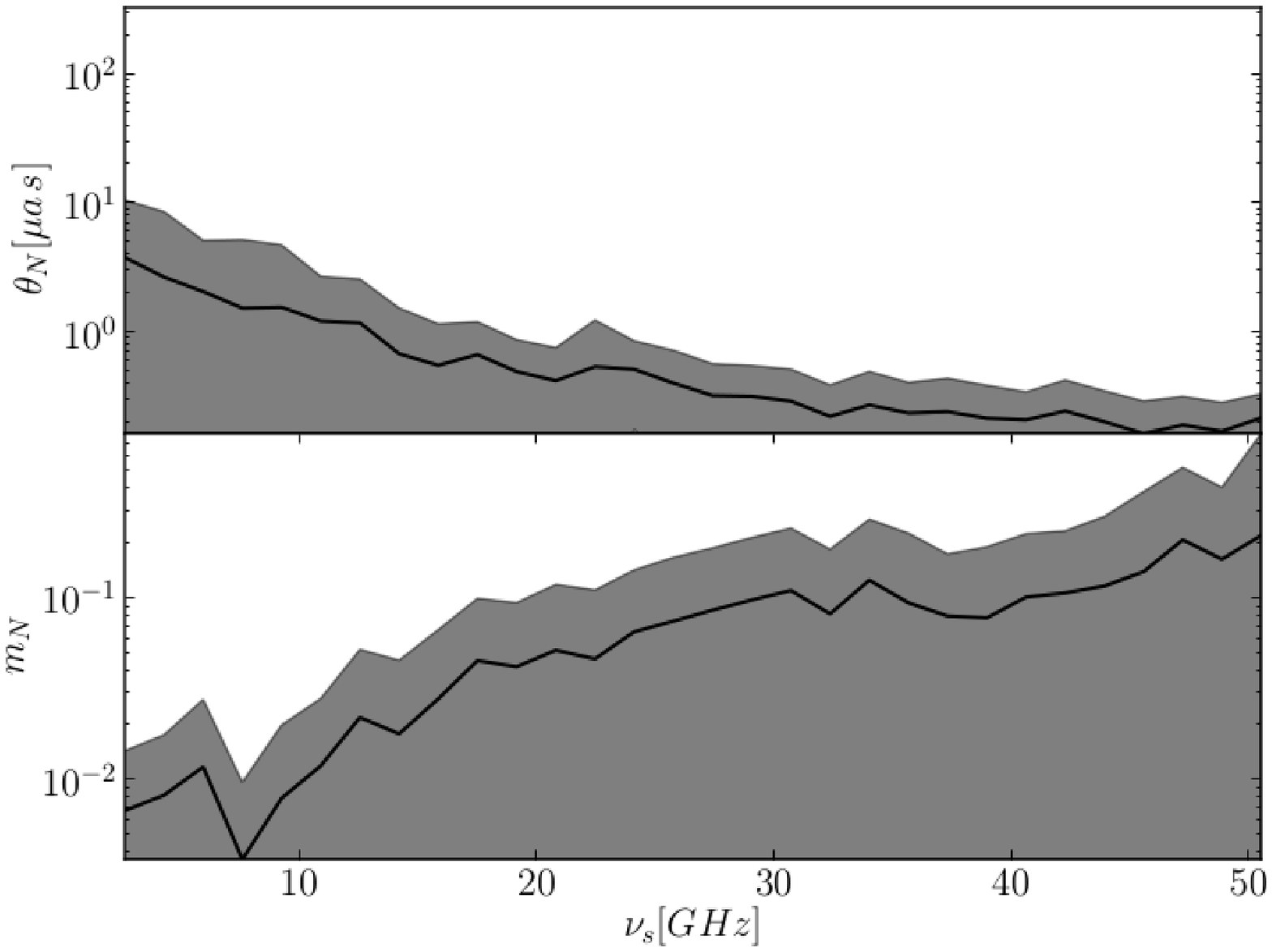}
\fi
\end{center}
\caption{IGM refraction angle (upper panel) and modulation index (lower) as a function of frequency for a source located at $z=2$; the solid lines indicate the mean and the shaded regions indicate 1-$\sigma$ fluctuations. 
\label{IGM_source_overview_fig}}
\end{figure}
Scintillation depends on frequency and therefore it is instructive to isolate such dependence in our results. To this aim we consider again a source at $z=2$ and allow its restframe frequency to vary in the range $[5,100]$ GHz. Contrary to the ISM case (Sec. \ref{ref_ISM_scintillation}) the source is approaching the strong scintillation regime ($\theta_{F}\ll\theta_{d}$) as the frequency increase. Thus, the frequency dependence of both $\theta_N$ and $m_N$, shown in Fig. \ref{IGM_source_overview_fig} closely follows the analytical predictions of eqs. \ref{angolodiff} and \ref{m_r-goodman}: $\theta_{N}\propto\nu_{s}^{-11/5}$, $m_{N}\propto \nu_{s}^{{17/30}}$.

\begin{figure}
\begin{center}
\ifpdf
  \includegraphics[width=8.3cm]{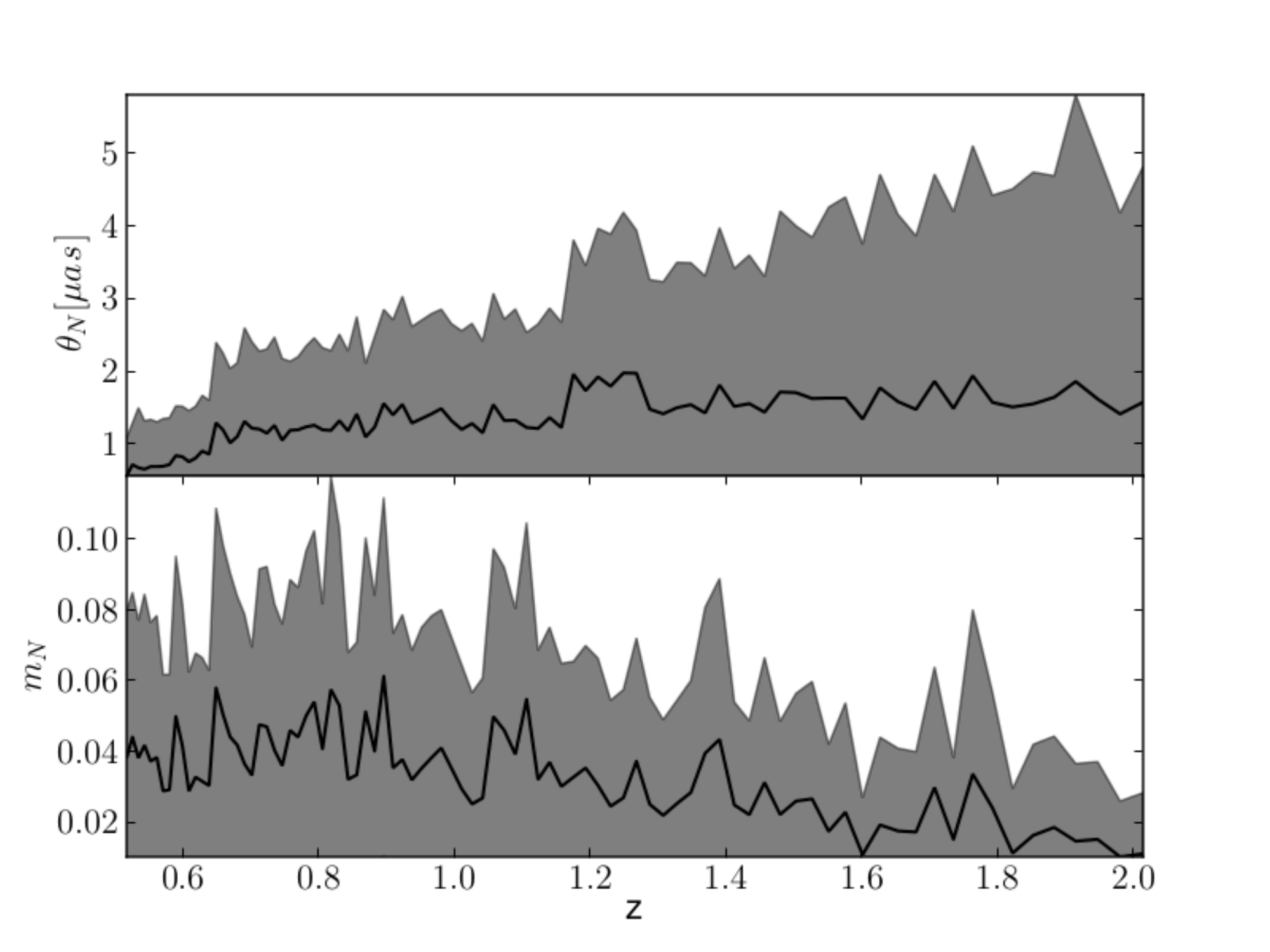}
\else
  \includegraphics[width=8.3cm]{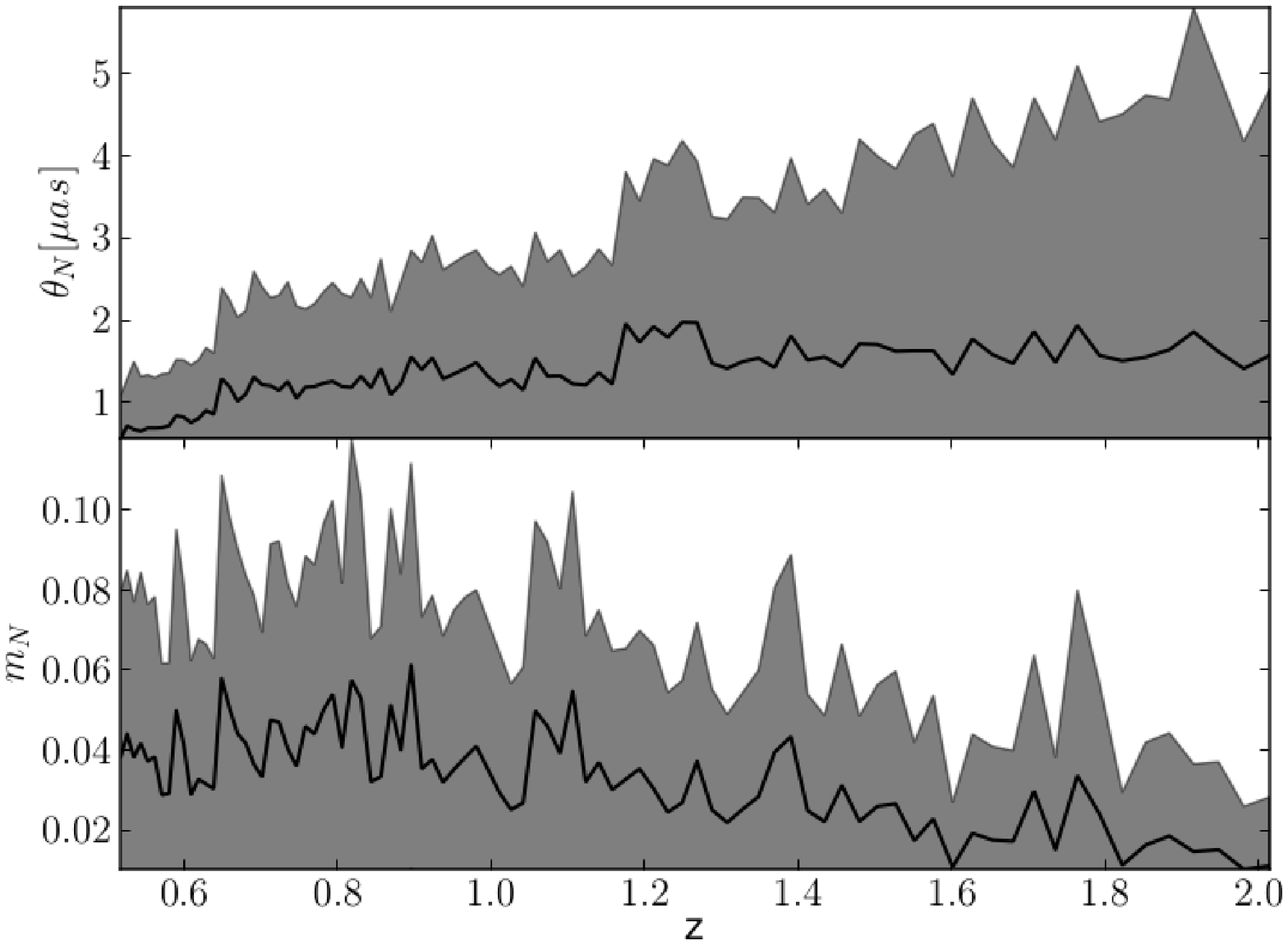}
\fi
\end{center}
\caption{IGM refraction angle (upper panel) and modulation index (lower)  as a function source redshift at $\nu_{s}=8$ GHz (restframe); the solid line indicate the mean and the shaded regions are the 1-$\sigma$ fluctuations.
\label{IGM_zeta_overview_fig}}
\end{figure}

Comparing the ISM results to the $m_N$ increasing trend for the IGM implies the existence of a critical frequency ($\nu_{s}\approx 30$) for which the the IGM modulation index becomes equal to the ISM one. 
Note that, for any given frequency, there is a negligible dependence on the source redshift as illustrated by Fig. \ref{IGM_zeta_overview_fig}.
\section{Comparison with observations}\label{ref_sezione_comparison}
The above results hint at the possibility that IGM scintillation as been so far largely underestimated. Recently, new high quality data from the Micro Arcsecond Scintillation Induced Variability (MASIV)~\citep[e.g.][]{Lovell:2008} survey have become available and allow a direct comparison with our results. MASIV has monitored during 4 observation epochs 482 quasars with $0\leq z\leq4$ at $\nu_{\obs}=5$ GHz. The variance in each ``light curve'' is characterized by a function of the time lag, taken to be $T_{\obs}=2$~d, defined as
\begin{equation}
	D(T_{\obs})=N_{t}^{-1}\sum_{i,j}\left(S_{i}-S_{j}\right)^{2}\, ,
\end{equation}
where $S_i$ is the flux density measurement of the $i$-th epoch normalized by the mean flux density of the source over all 4 epochs, $1\leq i,j\leq 4$ and $N_t$ is the number of pairs of flux densities; note that this observable quantity is directly related to the modulation index. The measure of $D(T_{\obs})$, in turn, allows to infer the source angular broadening \citep[i.e.][]{Lazio:2008}.

\citet{2007mru..confE..46R} pointed out a discrepancy between the data and the expectations from ISS theory; as a tentative explanation, \citet{Lazio:2008} suggested an intrinsic variability of the source. As an alternative \citet{2007mru..confE..46R} considered scintillation from a simplified IGM model, based on the $\HI$ column density distribution of Ly$\alpha$ forest absorbers. Such distribution is then translated into an electron density by assuming ionization equilibrium with a Haardt-Madau UV background \citep{Haardt:1996}. They concluded that IGM scintillation in such simple model cannot produce the relatively large ($\approx 10\ \mu$as) observed refraction angles, which cannot be explained by standard ISS theory. Note however that in our simulation such angular values are in the 1-$\sigma$ range of the $\theta_N$ for $\nu_s\lsim10$ GHz (see upper panel of Fig. \ref{IGM_source_overview_fig}).

Here we perform a comparison with the recent data reported by~\citet{Koay:2012}, which consist in a high redshift sub-sample of the MASIV survey. In this paper, the authors report the observation of 140 sources over a period of 11 days using VLA. Each observation lasted 1 minute with $\sim2$ hour intervals among them. The observations were done simultaneously at $\nu_{\obs}=4.9$~GHz and $\nu_{\obs}=8.4$~GHz. As in \citet{Lovell:2008}, \citet{Koay:2012} used $D(T_{\obs})$ to characterize the modulation.

We choose the emitting frequency $\nu_{s}=(1+z)\nu_{\obs}$ and we concentrate on the high frequency portion of the data. This choice is motivated by the fact that at higher $\nu_{s}$ the IGM contribution dominate over the ISS one (see Fig.~\ref{ISM_source_overview_fig} and Fig.~\ref{IGM_source_overview_fig}). Moreover, all observed sources are at high galactic latitude (see Fig. $12$ in \citet{Koay:2012}), thus further minimizing the ISS contribution. For this reason we do not include ISS in the subsequent analysis. We focus our attention on the $0.3\leq z\leq2$ range, containing 56 observed quasars.

\begin{figure}
\begin{center}
\ifpdf
  \includegraphics[width=8.3cm]{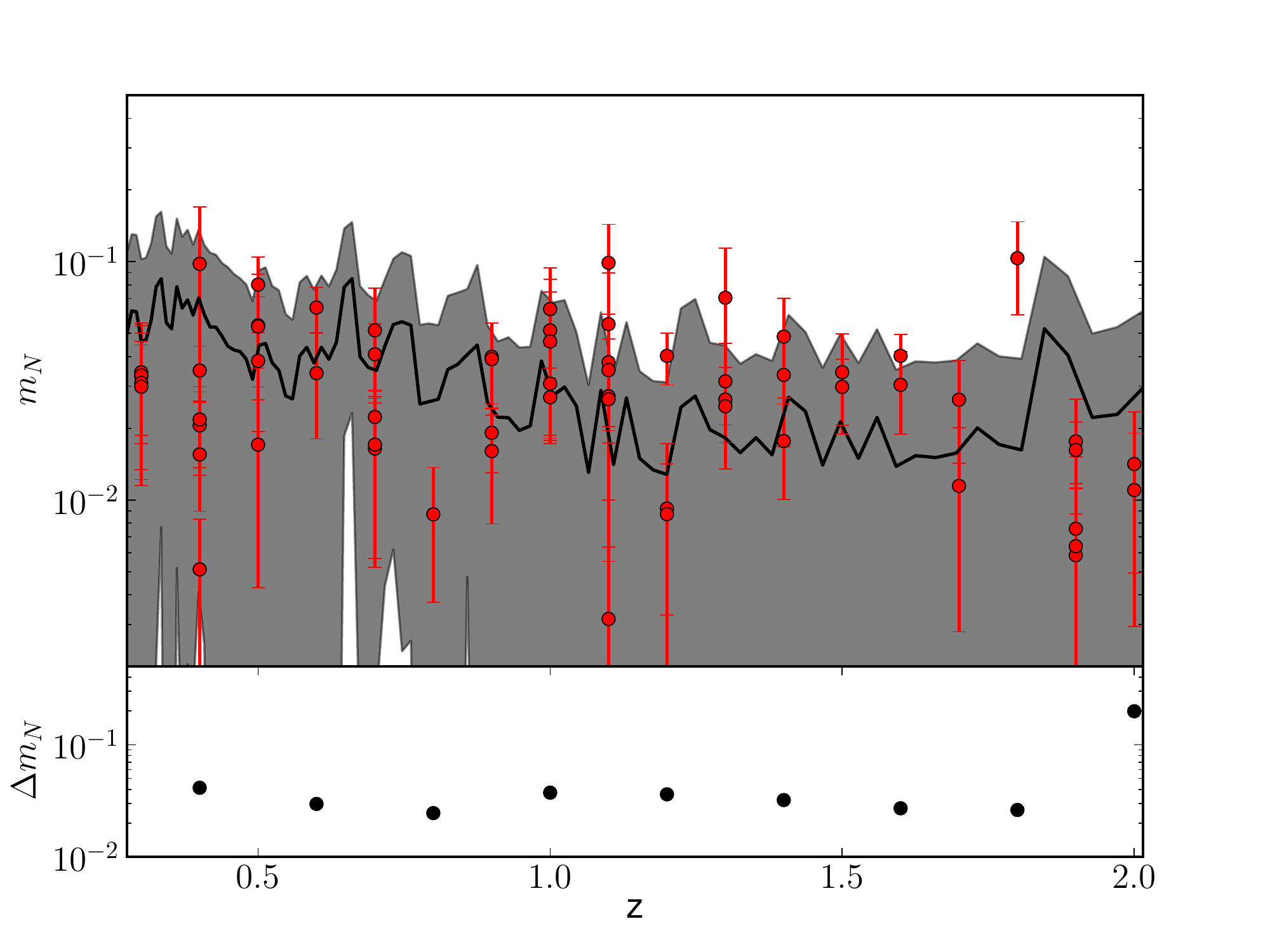}
\else
  \includegraphics[width=8.3cm]{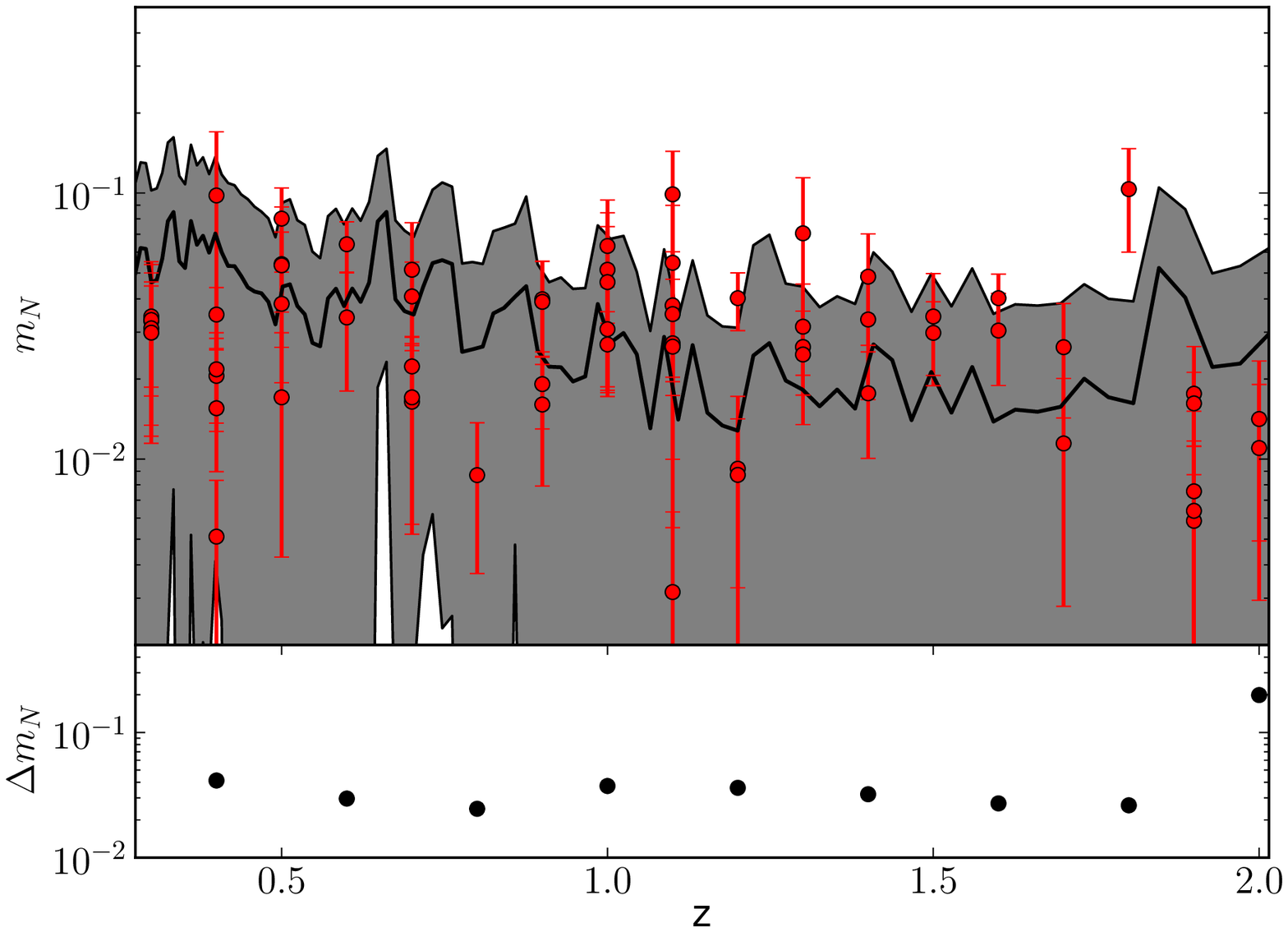}
\fi
\end{center}
\caption{
\textit{Upper panel:} IGM modulation index for sources at different redshift and restframe frequency $\nu_{s}=(1+z)\nu_{\obs}$ with $\nu_{\obs}=8.4\mbox{ GHz}$. The solid black line is the mean while the shaded region indicates the 1-$\sigma$ fluctuation; data (red points) and their $95\%$ confidence levels (solid red errorbars) are taken from \citet{Koay:2012}. \textit{Lower panel:} Fractional residuals after data subtraction from the model (as defined by eq. \ref{definizione_DeltamN}).
\label{IGM_zeta_frequenza_osservativo_fig}}
\end{figure}
The detailed comparison between data and our model is shown in Fig.~\ref{IGM_zeta_frequenza_osservativo_fig}, along with the fractional residuals after data ($D$) subtraction from the model ($M$):
\begin{equation}\label{definizione_DeltamN}
	\Delta m_{N}(z)= \left| \frac{ \left\langle m_{N}(M,z,\Delta z)\right\rangle -\left\langle m_{N}(D,z,\Delta z)\right\rangle }{ \left\langle m_{N}(D,z,\Delta z)\right\rangle } \right|;
\end{equation}
here the bracket operator stands for average on redshift bins of size $\Delta z=0.2$, which for $m_{N}(D)$ corresponds to averaging over $\sim 7$ data points per bin. Note that the mean residual is $\langle\Delta m_{N}\rangle=0.0486$, with little dependence on the bin size. Thus we can confidently state that our model correctly reproduces the data, allowing us to interpret the intensity modulation of these high latitude sources in the $[0.3,2]$ redshift range as being almost entirely due to intergalactic scintillation. This is a remarkable result in the light of the failure of previous interpretations based on the assumption of a smooth IGM. 

Moreover, there are three sources having modulation indexes outside the 1-$\sigma$ range of the model. This could be explained by the presence of a cluster in the l.o.s.. Additional measurement and a dedicated analysis may confirm this interpretation.
\section{Conclusions}\label{ref_sezione_conclusioni}
In this paper we have presented a new scheme to compute the intergalactic refractive scintillation of distant sources in combination with AMR cosmological simulations. Such scheme is physically motivated by the interpretation of scintillation as a Levy process \citep[e.g.][]{Boldyrev:2003,Boldyrev:2006}, and represents an extension of the thin screen approximation \citep[i.a.][]{Rickett:1990,Goodman:1997,Ferrara:2001}, suitable to treat both interstellar and intergalactic scintillation. Before applying our model to the IGM we have successfully validated our scheme using ISS data recently collected by \citet{Rickett:2006}.

To model the IGM we have performed extensive numerical simulations based on the public cosmological code {\tt RAMSES} \citep[i.e.][]{Teyssier:2002}. By assuming a $\Lambda$CMD cosmology and an external UV ionizing background we have followed structure formation up to $z=0$ in a 100 $\,h^{-1}$ Mpc box at high resolution. 

We have applied our scheme to the obtained cosmological density field and derived the scintillation signal induced by the IGM on a source emitting at frequency $\nu_s$ (in the range $5\leq\nu_s\slash\mbox{GHz}\leq50$) and redshift $z$ ($0.3\leq z\leq2$). By comparing the scintillation resulting from the cosmological simulation and a log-normal model for the IGM density fluctuations, we have isolated the contribution from largely non-linear overdensities dominating the extreme values of the modulation index. Finally we have compared our theoretical predictions to the experimental results obtained by \citet{Koay:2012} for an intermediate redshift subsample of the MASIV survey. The main results can be summarized as follows.
\begin{description}
 \item[(a)] The simulated IGM equivalent scattering measure, when averaged over $2\times 10^{4}$ l.o.s., is $\langle\mbox{SM}_{\equ}\rangle=3.879$, i.e. almost 40 times larger than expected from a smooth IGM. This value is also about 3 times larger than obtained assuming that the IGM overdensities can be described by a log-normal distribution. This outlines the importance of a correct description of the density field to compute scintillation effects.
 \item[(b)] For a source located at $z=2$ the mean refraction angle at $\nu_{s}=5$ GHz is $\langle\theta_N\rangle=1.77 \pm 0.11\,\mu$as with an r.m.s. of $\sigma=5.58 \pm 1.95\,\mu$as. Note that $\sim2\%$ of the l.o.s. have $\theta_N$ values as large as $70\,\mu$as. This result is important to interpret MASIV observations that require refraction angles as large as $\sim10\,\mu$as to explain the data in terms of IGM scintillation. As this values cannot be obtain by smooth IGM density models, \citet{2007mru..confE..46R} alternatively suggested that the source should be intrinsically variable. Our result show instead that IGM scintillation can provide an alternative viable solution.
 \item[(c)] For a $z=2$ source the average modulation index ranges from 0.01 ($\nu_s=5$ GHz) up to 0.2 ($\nu_s=50$ GHz). For $\nu_s>30$ GHz the IGM contribution dominates over ISS modulation, and scintillation can be used as a probe of IGM physics.
 \item[(d)] We analyze the observations from \citet{Koay:2012}, which are performed at $\nu_{\obs}=8.4$ GHz for sources in $0.3\leq z\leq2$. Because the high galactic latitude and emitting frequency ($10.92\leq \nu_s \leq 25.2$) of the sources, we can neglect the Galatic ISM contribution. Within our model the IGM produces a modulation index consistent within $4\%$ of \citet{Koay:2012} observations. This implies that for this sub-sample of objects the observed scintillation can be explained by IGM scintillation alone.
\end{description}

Scintillation as a tool to probe the ionized IGM is attractive due to its strong ($\rho^2$) density dependence and insensitivity to temperature thus allowing to trace both the cool and the warm diffuse components. The backdraw is that, being an integral quantity, it cannot yield precise spatial information of the underlying density structure along the line of sight. Moreover, inverting the measured $\theta_N$/$m_N$ to derive SM$_{\equ}$ is not simple. Nonetheless we have shown that for $\nu_s>30$ GHz the IGM scintillation dominates over the ISS, yielding average $m_N$ values in the observable range. Very high $m_N$ values are due to the presence of highly non-linear structures along the l.o.s., and we have given a method to calculate the relative probability of finding such events in an unbiased survey. Pinpointing such events in an observation can be used to infer the presence of large mass concentrations, like groups and clusters, possibly up to very high redshifts. In this case, our model offers a novel way to discover and study these objects. In the redshift range surveyed by~\citet{Koay:2012} ($0\leq z\leq4$), a significant modulation suppression for sources at $z\gsim2$ has been observed. In the future, we aim at extending our model to higher redshifts, enabling us to test whether this effect is due to evolution in the properties of the ionized IGM, as hinted by our analysis (see Fig.~\ref{IGM_zeta_overview_fig}).

The power of scintillation experiments to study large scale structures would be additionally boosted by the combination with an independent IGM probe, as for example the Compton-$y$ all-sky map, recently obtained by the \textit{Planck} satellite \citep{PLANCKXXI}.  These maps show an obvious galaxy cluster tSZ signal that is well matched with blindly detected clusters in the \textit{Planck} SZ catalogue. 
A joint study could be used to break the degeneracies affecting both techniques: since the scintillation amplitude depends on $n_e^2$, while the tSZ depends on both $n_e$ and $T$ it would be possible to infer $n_e$ and $T$ at the same time. In addition could also obtain important information on the IGM clumping factor $C=\langle n_e^2 \rangle/\langle n_e \rangle^2$. 

Alternatively, scintillation data can be combined with absorption line experiments which give detailed information on the neutral density fraction along the l.o.s. to reconstruct a the full (ionized+neutral) ionization field. 

Finally, and along similar lines, scintillation can be coupled with Faraday rotation studies \citep{Haverkorn:2013arXiv}. The magnitude of the effects is given respectively by the scattering and the rotation measure ($RM\propto\int n_{e} B_{\parallel}\mbox{d}s$, where $B_{\parallel}$ is the magnetic field component parallel to the l.o.s.). Due to the different density dependence one could get useful insights on the turbulent and magnetic field structure of the IGM. These experiments suitably fit the core science of forthcoming radio facilities as SKA.

\section*{Acknowledgments}
We thank K. Koay for  providing the data. CE acknowledges support from the Helmholtz Alliance for Astroparticle Physics funded by the Initiative and Networking Fund of the Helmholtz Association.
\bibliographystyle{mn2e}
\bibliography{scintillation}
\appendix
\section{Probability Distribution Functions and their errors}\label{app_PDF}
The PDFs in this paper are calculated with a kernel density estimate method, a technique similar that implemented in SPH \citep[e.g.][]{Monaghan:1992}. In general, for a scalar function $f$ whose values are known on the finite ensemble $\{x_{i}\}=\{x_1,\dots, x_M\}$, the density estimator on the $x$ interval is given by
\begin{equation}\label{equazionedensityestimator}
	f_{est}\left(x\right)=\sum_{i=1}^{M} f\left(x_{i}\right)K\left(x_{i},x\right)\, ,
\end{equation}
where $K$ is a Gaussian smoothing kernel:
\begin{equation}
	K\left(x_{i},x_{j}\right)=\left(2\pi h\right)^{-3\slash2}\exp\left[-\left(\frac{x_{i}-x_{j}}{h}\right)^{2}\right]\, ,
\end{equation}
with a constant bandwidth $h$ that we have adopted from \citet{Silverman:1986}
\begin{equation}
h=1.06\, \sqrt{\langle x_{i}^{2}\rangle} M^{-1\slash5}\, ;
\end{equation}
$\langle x_{i}^{2}\rangle$ is the variance of the sample. Note that $h$ represents the scale above which $f_{est}$ becomes a good approximation of $f$.

Let us take the following functional form for $f$
\begin{equation*}
f=
\left\{\begin{aligned}
&w(x)\quad&x\in\{x_{i}\}\\
&\ 0 &\mbox{otherwise}
\end{aligned}\right.
\end{equation*}
with $w(x)$ a proper weighting function, which, when not explicitly stated, we have set to a constant. This choice of $f$ enables us to interpret $f_{est}$ as a PDF
\begin{equation}
	\frac{\mbox{d}P}{\mbox{d}x}\left(x\right)\propto\sum_{i=1}^{M} w(x_{i}) K\left(x_{i},x\right)\, ,
\end{equation}
where the proportionality constant is obtained by normalization of the PDF. As usual the CDF is obtained by integrating the PDF
\begin{equation}\label{eq_pdf_appendice}
	P(<x)=\int^{x}_{\min_x}\frac{\mbox{d}P}{\mbox{d}x}(y)\,\mbox{d}y\, ,
\end{equation}
where $\min_x$ indicates the minimum of the support of the PDF.

To propagate errors in the initial sample, we use a bootstrapping method. Let $\{\epsilon_{i}\}$ be the relative errors associated with the sample $\{x_{i}\}$. From these two sets it is possible to construct the set $\{y_{i}\}$ defined by
\begin{equation}
	y_i =x_i\left(1+\epsilon_i\mathcal{R}\right)\, ,
\end{equation}
where $\mathcal{R}$ is a random variable, uniformly distributed in the interval [-1,1].

We label a particular realization as the set $\{y_{i}\}(\gamma)$. Using eq. \ref{eq_pdf_appendice} it is possible to calculate the associated PDF, $g(x,\gamma)$. Finally taking $N_{\pdf}$ realizations we can write the estimate for the PDF as an average on $\gamma$
\begin{equation}
	\frac{\mbox{d}P}{\mbox{d}x}(x)=\langle g(x,\gamma)\rangle\, ,
\end{equation}
with associated error given by the r.m.s. of the realizations
\begin{equation}
	\Delta\left(\mbox{d}P\slash\mbox{d}x\right)(x)=\sqrt{\left\langle\left[g(x,\gamma)-\left\langle g(x,\gamma)\right\rangle\right]^{2}\right\rangle}\, .
\end{equation}
To assure a suitable convergence through the paper we use $N_{\pdf}=5\times 10^{3}$.

\bsp
\label{lastpage}
\end{document}